\documentclass[11pt]{amsart}
\usepackage[english]{babel}
\usepackage{hhline,float,tabularx, multirow, amsmath, amstext,amssymb}
\usepackage{amsthm}
\usepackage{thmtools,thm-restate}
\usepackage{float}
\usepackage{marvosym}
\usepackage{mathpazo}
\usepackage{color,colortbl}
\definecolor{blue1}{RGB}{25,25,112}
\usepackage{graphicx}
\usepackage{caption}
\usepackage{subcaption}
\usepackage[foot]{amsaddr}
\usepackage{mathrsfs}
\usepackage{refcount}
\usepackage{latexsym}
\usepackage{stmaryrd}
\usepackage[margin=1in]{geometry}
\usepackage{cancel}
\usepackage{setspace}
\usepackage{bbm}
\usepackage{url}
\usepackage{hyperref}
\hypersetup{bookmarksnumbered=true,
	colorlinks=true,
	citecolor=blue1,urlcolor=blue1,linkcolor=blue1,
}

\usepackage{natbib,har2nat}
\usepackage{tikz} 
\usepackage{pgfplots}

\usepackage[colorinlistoftodos,prependcaption,textsize=small]{todonotes}
\presetkeys{todonotes}{disable}{}

\allowdisplaybreaks

\begin{document}
	
	\makeatletter
	\renewcommand\section{\@startsection{section}{2}%
		\z@{.7\linespacing\@plus\linespacing}{.5\linespacing}%
		{\scshape\centering}}
	\renewcommand\subsection{\@startsection{subsection}{2}%
		\z@{.7\linespacing\@plus\linespacing}{.5\linespacing}%
		{\normalfont\itshape\centering}}
	\makeatother
	\newtheorem{theorem}{Theorem}
	\newtheorem{observation}{Observation}
	\newtheorem{lemma}{Lemma}
	\newtheorem{proposition}{Proposition}
	\theoremstyle{definition}
	\newtheorem{definition}{Definition}
	\newtheorem{example}{Example}
	\newtheorem{corollary}{Corollary}
	\newcommand{\argmax}{\operatornamewithlimits{argmax}}	
	\newcommand{\argmin}{\operatornamewithlimits{argmin}}
	
	\newcommand{\bbE}{\mathbb{E}}
	\newcommand{\indic}{\mathbbm{1}}
	\newcommand{\umu}{\underline{\lambda}}
	\newcommand{\omu}{\overline{\lambda}}
	
	\newcommand{\rahulnote}[2][1=]{\todo[linecolor=red,backgroundcolor=red!25,bordercolor=red,#1]{#2}}
	\newcommand{\mattnote}[2][1=]{\todo[linecolor=green,backgroundcolor=green!25,bordercolor=green,#1]{#2}}
	\newcommand{\heskinote}[2][1=]{\todo[linecolor=blue,backgroundcolor=blue!25,bordercolor=blue,#1]{#2}}
	
	\title{Selling Certification, Content Moderation, and Attention}
	
	\author[Bar-Isaac]{Heski Bar-Isaac$^{\dag}$}
	\author[Deb]{Rahul Deb$^{\between}$}
	\author[Mitchell]{Matthew Mitchell$^{\ddag}$}
	
	\date{\today}
	
	\thanks{$^\dag$\href{mailto:heski.bar-isaac@rotman.utoronto.ca}{\nolinkurl{heski.bar-isaac@rotman.utoronto.ca}}; Rotman School of Management, University of Toronto \\ $^{\between}$\href{mailto:rahul.deb@bc.edu}{\nolinkurl{rahul.deb@bc.edu}}; Department of Economics, Boston College \\ $^\ddag$\href{mailto:matthew.mitchell@rotman.utoronto.ca}{\nolinkurl{matthew.mitchell@rotman.utoronto.ca}}; Rotman School of Management, University of Toronto \\ We would like to thank Mark Armstrong, Itay Fainmesser, Kinshuk Jerath, Byung-Cheol Kim, Marco Ottaviani, Yossi Spiegel, Greg Taylor, and participants at numerous conferences and seminars for their insightful comments.
	}
	
	\begin{abstract}
	We introduce a model of content moderation for sale, where a platform can channel attention in two ways: direct steering that makes content visible to consumers and certification that controls what consumers know about the content. The platform optimally price discriminates using both instruments. Content from higher willingness-to-pay providers enjoys higher quality certification and more views. The platform cross-subsidizes content: the same certificate is assigned to content from low willingness-to-pay providers that appeals to consumers and content from higher willingness-to-pay providers that does not. Cross-subsidization can benefit consumers by making content more diverse; regulation enforcing accurate certification may be harmful.
	\end{abstract}
	
	%
	
	\maketitle
	
	\mattnote[inline]{decide carefully on where to use R, including the appendix}
	
	\section{Introduction}
	The digital environment is overwhelming. There are over a billion websites on the internet, Tiktok hosts billions of videos, more than two trillion posts have been created on Facebook.\footnote{These estimates are drawn from \url{https://siteefy.com/how-many-websites-are-there/}, \url{https://www.usesignhouse.com/blog/tiktok-stats}, and \url{https://www.wordstream.com/blog/ws/2017/11/07/facebook-statistics} all accessed on June 7, 2024.} In such an environment, it is little surprise that attention is a key resource and the platforms that control access to such attention can be increasingly sophisticated and profitable in doing so. This has not escaped economists and by now there are surveys on both digital economics \cite{goldfarb2019digital} and social media \cite{aridor2024economics}. Meanwhile regulators have become concerned over how platforms, possibly with monopoly power, might manage attention in ways that favor profits over consumer welfare.
	
	In this paper, we examine how a monopoly platform maximizes profits by governing and charging content providers for two key aspects of attention: the frequency with which posts gets viewed (steering) and how posts are presented to viewers (certification). We broadly refer to these two channels as content moderation. We show that imperfect certification, where the quality differences in content that matter to consumers are obscured, can increase content diversity when views are for sale. As a result, consumers can, but need not, benefit from certification for sale relative to enforced perfect certification.  This informs recent policy discussion around permitting the sale of certification on platforms.
	
	We analyze a model where a platform sells views and certification to content providers interested in consumers' attention and engagement. The platform can observe whether content providers are good or bad; content from the former is valued by consumers and content from the latter is not. Therefore, from the consumers' perspective, the platform has the perfect ability (but possibly not the incentive) to screen content.  Consumers' attention is governed by their intrinsic interest (which makes it ever harder to generate engagement) and by their expectations that any piece of content will be something they value seeing. These beliefs are governed by an understanding of the likelihood that content presented in a given way (or, equivalently, with a given certificate) arises from a good provider. In turn, these beliefs govern consumers' attention which is what content providers—both good and bad—value.  We focus on the application of an online platform selling steering and certification because it corresponds to an important recent development in these markets, and in their regulation. However, in the conclusion section, we point to a wider set of applications that are captured by our framework.
	
	Content providers want attention: to be seen and, hopefully, trusted so that their content is engaged with. Being seen is necessary to command attention and get engagement but it is not sufficient. In addition to steering content directly, the platform can attach an arbitrary message (the certificate) in order to provide consumers with information about quality (the likelihood that the content is from a good provider).\footnote{In practice, this may, most obviously, include ``checkmarks" for verified status as Twitter began to charge for in November 2022. But it could also include position on a page, aspects of display, ancillary information such as ``your great aunt Naima likes this post," etc.} This corresponds to everything the consumer sees before deciding whether to further inspect the content. The platform cannot tell the willingness to pay for engagement that good providers have, and it price discriminates by offering different bundles of views and messages. Conversely, we assume that bad providers have a known willingness to pay for attention.
	
	One might suppose that the ability to control views alone is sufficient to exercise full monopoly power. We show that this is not the case. The ability to vary (and charge for) certification can generate additional platform profit notably through imperfect certification (whereby content from both good and bad providers are assigned the same certificate). This allows for revenue from content that consumers would prefer not to see. In turn, one might suppose that consumers suffer from a platform's ability to sell certification as well as views. We show that there is an economic force acting against this intuition that can overwhelm it. Specifically, consumers might gain from seeing good content whose provider might have relatively little value from receiving this attention. Imperfect certification can, in effect, subsidize views of such content in order to to sell views to bad content providers and so can improve content diversity. 
	
	To see why content diversity is impacted by imperfect certification, start from the case of perfect certification. This implies that the platform is mandated to use its certification technology to ensure that bad providers receive no views (one can imagine bad content is clearly labeled as such) and consequently, no attention. Under perfect certification, our platform's profit maximization problem of choosing how many views to provide to each content provider and at what price reduces to a classic model of second-degree price discrimination equivalent to the seminal analysis of \citet{mussa1978}. Since good content providers vary in their private valuation for engagement, those that value it more, pay more and receive more views. Lower willingness-to-pay providers get fewer views and some providers with a positive willingness-to-pay are excluded entirely.
	
	Now consider the case where there is only one imperfect certificate available, and its quality is fixed. This is equivalent to all content receiving the same message but the aggregate share of good content being known. As a result, for each view that good providers receive, bad providers receive views in some fixed proportion. In this case, views from good providers allow the platform to generate revenue from bad providers. This allows the platform to profit from selling views to lower value good content providers that would be excluded under perfect certification.
	
	In the fully profit maximizing choice of views and certification, higher willingness-to-pay content providers receive both more views and a message that makes their content more trusted. High enough willingness-to-pay good content providers get perfect certification. Lower willingness-to-pay good providers are not excluded but instead assigned lower quality certification. As we argue, this expansion of content diversity (relative to perfect certification) might benefit consumers by making the platform more egalitarian. 
	
	To understand why platforms might have changed their approaches to selling attention — notably Twitter's move to charge for ``verified status" in November 2022 — we examine comparative statics in the model. In particular, we highlight how a reduction in what platforms can charge for ads can lead to a move away from perfect certification and that cheaper targeting does not affect the quality of certification. In addition, we highlight how the nature of attention affects whether or not platforms engage in imperfect certification. Specifically, we vary the convexity of attention as a function of content quality with the aim of capturing the extent to which consumers are put off by bad content.
	
	Both steering and certification by platforms have come under regulatory scrutiny more generally. The importance of views is central to algorithmic design around ranking which is under increasing regulatory scrutiny, as in \citet{CMAalgorithms}. The recognition that the presentation of some information might affect the extent to which it is deemed worthy of attention is, of course, at the heart of disclosure regulation (in the context of the kind of social media application that inspires our study, see \citet{mitchell2021free}, for example) and central to understanding the discussion around disclosure and certification (\citet{dranove2010quality}  provides an excellent overview). Content certification for sale has become an issue as well, with the European Commission announcing that they will seek remedies against X for its practice of selling certification through checkmarks.\footnote{See \url{https://ec.europa.eu/commission/presscorner/detail/en/ip_24_3761}} 
	
	Our results directly impact the policy debate around certification for sale. While the optimal mechanism always results in weakly greater content diversity relative to perfect certification, consumers are faced with more bad content. Using total engagement as a measure for consumer welfare, we provide conditions under which enforcing perfect certification does, and does not, benefit consumers. We thus show that there is no single correct answer to the question of whether the sale of certification should be prohibited.
	
	The paper is organized as follows. In the next section, we review the literature. Then, we present the model and study several benchmarks: an engagement maximizing planner, followed by profit maximization when restricted to simple certification with only one or two certificates. The one certificate case includes perfect certification as a special case. We then derive the optimal mechanism and develop comparative statics. Since it is an important policy concern, we study the comparison of the optimal mechanism to enforced perfect certification in detail. Throughout, we illustrate the central forces through an example where attention is a linear function, which allows for an explicit and graphical illustration. Finally, via a simple extension, we show that our framework can accommodate bad content that causes damage and social media addiction, and highlight the intuitive effects of both.
	
	\subsection{Related Literature}
	
	In considering a platform that sells attention both through certification and more prominent views, we bring together literatures that have considered each of these aspects separately.
	
	\subsubsection{Platform Steering without Certification}
	
	Our approach is based on solving a second-degree price discrimination problem (in the style of \citet{mussa1978}). Others have considered that the two-sided nature of platforms changes the standard analysis when the platform collects revenue from both sides. Papers include \citet{choi2015net}, \citet{bohme2016second}, and \citet{jeon2022second}. 
	
	\citet{choi2015net} and \citet{bohme2016second} explore how platforms can maximize profits by differentiating prices for different sellers or users, which indirectly steers users. \citet{jeon2022second} further expand on this by investigating second-degree price discrimination in monopoly platforms. Focusing on the impact of platform steering on sellers' incentives, \citet{Johnsonsteering} and \citet{ichihashi2023buyeroptimal} argue that steering alters sellers' competitive behavior. By influencing the exposure of sellers' products to consumers, platforms shape the strategic decisions that sellers make. Our model focuses on the social media context, where content is not priced directly by providers to consumers and can be steered to many consumers at low cost. Moreover, the platform sells not only ``quantity'' in the form of views but also quality (via certification), which affects consumer attention. It is precisely the interaction between these two that is the focus of our analysis.
	
	In the social media context, various papers have studied steering without certification or direct monetary payment for steering. These papers focus on how platforms recommend different types of content to viewers, and how they incentivize content creation. Examples, building on \citet{ghosh2011incentivizing}, include \citet{bhargava2022creator}, \citet{srinivasan2023paying}, 
	\citet{qian2024digital}, and \citet{chen2025algorithmic}.
	
	\subsubsection{Certification without Steering}
	
	\citet{dranove2010quality} provide a wide-ranging survey of the literature on certification. In this literature, \citet{lizzeri1999information} is an early contribution that shares our conclusion that imperfect certification is profit-maximizing for the certifier as it allows good but not great sellers to charge a higher price.\footnote{This literature has developed in several ways. For example, \citet{ali2022sell} consider uninformed sellers who can conceal that they have been tested. Following the subprime crisis in 2007, a broad literature has considered certification in the credit-rating industry for which useful surveys can be found in \citet{white2010markets} and \citet{jeon2013credit}.} Among more recent papers, perhaps \citet{bouvard2018two} is the most related in very clearly highlighting that profit-maximizing certification trades off pooling different types of sellers to earn more from low-quality sellers but diluting quality too much alienates consumers (which in our environment corresponds to receiving less attention). Our certification is mixed with direct steering, so that the certification need not do the steering on its own.
	
	\subsubsection{Indirect Steering through Search}
	
	Direct steering means that the platform in our model must consider the scarcity in the total possible attention (which we capture by a convex cost for the platform in finding relevant viewers for good content). Consequently, our analysis shares features with the literature that has focused on how platforms sell off scarce slots (including seminal contributions by \citet{edelman2007internet}, \citet{chen2011paid}, \citet{armstrong2011paying} and \citet{athey2011position}, or, more recently, \citet{bar2022monetizing} who contrast different sales mechanisms). We share with much of this literature the observation that given the mechanisms through which these positions are sold, consumers draw equilibrium inferences about the quality of offerings associated with their rankings (different certificates, in our work). 
	
	Our focus on how profit incentives might lead a platform away from perfect certification and efficiently allocating views is somewhat reminiscent of a literature that examines biased intermediaries and search diversion (\citet{de2014integration}, \citet{hagiu2014search}, \citet{burguet2015google}, and \citet{de2019model}), though much of this literature is more focused on the consumer search process.
	
	\subsubsection{Content Moderation (Not for Sale)}
	
	Content moderation without direct monetary incentives has become an increasingly important area of study as platforms attempt to balance openness with the need to manage harmful or misleading content. \citet{MadioQuinn2024} study a platform that manages the value of a third party (advertiser) interest in content moderation. \citet{zou2025designing} like this paper highlights a tradeoff between quality and variety but in an environment where entry and content quality are endogenous and affected by a recommendation system that aims to maximize consumer surplus and does not earn revenue from content providers. 
	
	\citet{KominersShapiro2024} explore a sender-receiver game where a platform can moderate content, in the sense of manipulating what is seen by the receiver for any message sent. This conforms to our idea of the general description of content moderation, but in a different modeling setting. \citet*{acemoglu2023content} study a model where content moderation is about sharing, and how content sharing between consumers might be regulated. Here content moderation is more easily thought of as relating to the content shared between users and not between providers and consumers. 
	
	Another form of content moderation that can be considered part of the certification process is disclosure regulation. Here content that is paid for must be combined with a message that indicates that this is the case. \citet{Inderst} examine a general model of disclosure regulation. \citet{mitchell2021free} and \citet{fainmesser2021market} model how disclosure regulations might impact relationships between content providers (``influencers'') and consumers (``followers''). \citet{ershov2024advertising} 
	provide evidence on the impact of this form of content moderation.\footnote{Papers on rules surrounding deceptive sales practices such as \citet{corts2013prohibitions}, \citet{corts2014finite}, \citet{glaeser2010regulating}, and \citet{rhodes2018false} also highlight the role that some form of rules on messages might play. This also fits our description of content moderation.}
	
	
	
	\section{Model and preliminaries}\label{sec:model}
	
	We study a price discrimination problem of a platform through which content providers reach consumers. We now describe the model in detail starting with the content providers.
	
	\paragraph{\emph{Content Providers.}} Content \textit{providers} can either be $g$ood or $b$ad; good content can be of value to interested consumers (described below), but bad content cannot. There is a continuum of a unit mass of good providers whose private \textit{values} $\theta\in [0,\overline{\theta}]=:\Theta$ are distributed according to $F\in\Delta(\Theta)$ that has a continuous, positive density $f(\cdot)>0$. $\theta$ captures the extent to which a good content provider values engagement. There is an unlimited mass of bad providers who all have the same value for their content being read.\footnote{Given our assumptions on the costs of the platform that follow, we could equivalently assume that there is a fixed mass of bad providers but that their content can be spread widely.}
	
	The amount of \textit{engagement} $a v_g$ with good content is the product of the number $v_g\in \mathbb{R}_+$ of \textit{interested} views that the platform provides a good content provider and the \textit{attention} $a\in[0,1]$ that these viewers pay to the content. The utility of a good content provider with value $\theta$ from a given level of engagement $a v_g$ is $\theta a v_g$.
	
	Consumers never engage with content from bad providers and thus they only value their content being read. A bad content provider receives utility $a v_b$ from $v_b\in \mathbb{R}_+$ views that pay a level $a\in[0,1]$ of attention to their content.
	
	There are three differences between good and bad providers. First, good providers care about engagement whereas bad providers only care about being read. This distinction is not yet apparent from the payoffs (since they both depend on the product of attention and views) but will become clear below when we define the platforms costs (in essence, it is more costly for the platform to provide interested views to the good providers). Second, there is no heterogeneity in the marginal valuations of the bad providers.\footnote{This assumption is purely for technical convenience since it avoids the complications that arise with multidimensional private information. It is also consistent with the assumption that there are infinitely many bad providers since the platform can sell to those with highest value.} Third, there is a limited mass of good providers but we assume (for realism) that there is an unlimited amount of bad content since, in particular, its generation can be automated. Alternatively, the assumption is consistent with the idea that, since bad content is never of interest to the consumer, a single piece could be reused and shown to many consumers.

	\paragraph{\emph{Targeting of Content and Platform Costs.}} The platform can distinguish between good and bad providers.\footnote{For example, platforms can distinguish between good and bad providers exactly as users can but need only do so once on behalf of all potential viewers. Further, in addition to observing content directly, platforms monitor consumer reactions and other measures of engagement.} However, the platform does not know good providers' valuations for engagement. The platform chooses the number of views to direct to each provider's content. Directing $v_b$ untargeted views at content from a bad provider costs the platform $\gamma v_b$ where $0<\gamma<\min\{\overline{\theta},1\}$.\footnote{This assumption ensures that it is not automatically unprofitable for the platform to serve either type of providers. If $\gamma>\overline{\theta}$, good providers cannot be served profitably. If $\gamma>1$, bad providers cannot be served profitably.} This opportunity cost reflects, for example, that the platform could direct advertisements at consumers instead of content. 
	
	The same opportunity cost is also present when directing views at content from good providers. However, these providers only value engaged users and this is the source of an additional cost: the platform needs to search for such users of whom there is a smaller pool.\footnote{Of course, the platform could send out untargeted views with the hope of randomly finding interested users. Suppose that the unconditional probability of a user being interested in an untargeted post is $\rho$. The optimal mechanism we derive using the cost function $c$ will remain optimal for a sufficiently low $\rho$. This is because it is cheaper to target than to pay the opportunity cost of the number of untargeted posts required to generate the same amount of interest.\label{footnote_rho}} The platform faces a cost $\gamma v_g + c(v_g)$ of providing $v_g$ interested views to content from a good provider, where $c:\mathbb{R}_+\to \mathbb{R}_+$ is a strictly  increasing, strictly convex and differentiable function that satisfies $c(0)=c'(0)=0$ and $\lim_{v_g\to\infty}c'(v_g)=\infty$. The convexity of $c$ corresponds to the increasing difficulty of targeting the content of a given good provider with interested viewers as the content is shown more times. Therefore, the convexity is at the level of a given content provider.
	
	\paragraph{\emph{Consumers.}}  There are a unit mass of consumers who may view content. Each piece of content is accompanied by a message $m$.\footnote{The message $m$ can be broadly understood as reflecting whatever the consumer uses to form beliefs about the quality of content before reading it and may include location on a page, information such as how many others ``liked'' a post, their identity, explicit checkmarks etc.} Consumers observe the message $m$ and decide whether to read the content to learn more it. Reading requires paying a 
	cost $q \geq 0$, distributed according to  strictly increasing, differentiable cumulative distribution function $A(q)$. The consumer receives a payoff of 1 if they decide to read the content and it turns out to be good content that matches their interest (such content generates engagement). They receive a payoff of 0 otherwise. In Section \ref{subset:losses} we generalize to negative payoffs for consumers from reading bad content.\footnote{Such a change might reflect that bad content includes scams and other costly outcomes. As will become apparent, allowing for heterogeneity in the extent of harm from bad content, rather than reading costs, would lead to similar effects.}
	
	Upon observing a message $m$, consumers assign probability $\lambda\in[0,1]$ to the content being good and of interest. The expected payoff from reading the content is therefore $\lambda-q$. Consequently, a consumer chooses to read the content if, and only if, $q\leq\lambda$. In other words, $A(\lambda)$ is the likelihood that content labeled with message $m$ is read by consumers. We henceforth refer to $A$ as the \textit{attention function}. Since engagement is the consumer's payoff, the engagement maximizing benchmark is a natural consumer welfare standard that we analyze below. Aside from allowing a welfare calculation, the details of the underlying interpretation of the attention function as a cost calculation embodied in $q$ is inessential. The same predictions arise, under any functional relationship between beliefs and attention that satisfies our assumptions (that is, a strictly increasing, differentiable, cumulative distribution function). These could represent other forms of heterogeneity across and within consumers that drives attention.
	
	\paragraph{\emph{Platform Pricing.}} The platform price discriminates by offering a (direct) \textit{mechanism} to providers. The mechanism consists of four functions. These correspond to the message or \textit{certificate} assigned to a good provider claiming to be of value $\theta$; the number of targeted views that this provider receives; the number of untargeted bad provider views that are assigned to value $\theta$'s certificate for each report of $\theta$; and the price that the good provider pays to receive the certificate. As discussed below, the surplus from bad providers is fully extracted and so we do not need incorporate their payments into the mechanism.
	
	These functions are written as follows: 
	\begin{align*}
		& M:\Theta\to \mathbb{R},\\
		& V_g:\Theta\to \mathbb{R}_+,\\
		& V_b:\Theta\to \mathbb{R}_+,\\
		& P:\Theta\to \mathbb{R}.
	\end{align*}
	
	Note that the only private information is the value of the good providers so the mechanism is a function of $\Theta$. A good provider whose value is $\theta$ pays $P(\theta)$ to receive a certificate $M(\theta)$ and $V_g(\theta)$ targeted views. Additionally, there are $V_b(\theta)$ untargeted views by bad providers that are also assigned certificated $M(\theta)$ for each report of $\theta$. 
	
	For every $m$ in the image of $M$, we use
	$$\Lambda(m) = \frac{\mathbb{E}\left[V_g(\theta) \mid M(\theta) = m\right]}{\mathbb{E}\left[V_g(\theta) + V_b(\theta) \mid M(\theta) = m\right]}, $$
	to denote the fraction of good views assigned to certificate $m$ or the \textit{quality} of the certificate for short. When both the numerator and denominator are zero in the above fraction, $\Lambda(m)$ can be chosen arbitrarily.
	
	The platform's mechanism design problem is
	\begin{align}\label{eq:prin_prob}
		\begin{split}
			& \max_{V_g,V_b,M,P} \int_{\Theta} \left[ P(\theta)+ A(\Lambda(M(\theta)))V_b(\theta) -\gamma(V_b(\theta)+V_g(\theta)) - c(V_g(\theta)) \right]f(\theta)d\theta, \\
			& \text{subject to}\\
			& \theta A(\Lambda(M(\theta)))V_g(\theta) - P(\theta) \geq \max\{\theta A(\Lambda(M(\theta')))V_g(\theta') - P(\theta'),0\}\qquad \text{ for all }\theta,\theta'\in \Theta.
		\end{split}
	\end{align}
	The objective function sums the total payments received by the platform net of the costs of providing the views to the content providers. Each good provider of value $\theta$ pays the platform $P(\theta)$. Bad providers do not have any private information so the platform simply charges them the utility that they receive from being assigned certificate $M(\theta)$ and receiving $V_b(\theta)$ views, which is $A(\Lambda(M(\theta)))V_b(\theta)$. $A(\Lambda(M(\theta)))$ is the fraction of consumers who pay attention to content marked with a certificate $M(\theta)$. The cost of $V_b(\theta)$ untargeted views is $\gamma V_b(\theta)$. The cost of $V_g(\theta)$ targeted views is $\gamma V_g(\theta) + c(V_g(\theta)))$ due to the additional component associated with targeting. The constraint captures both  incentive compatibility and individual rationality constraints for good providers.
	
	The case that $A(1)\leq\gamma$ trivially implies that no views will be directed to bad providers since the costs to the platform would then be higher than the value of the views. Therefore, we henceforth focus on the more interesting case $A(1)>\gamma$ and we normalize $A(1)=1$ (recall that $A(0)=0$). 
	
	Before we begin analyzing the above problem, a few comments are in order. First, observe that because providers are infinitesimal, a misreport by value $\theta$ as a value $\theta'\neq \theta$ does not affect the quality $\Lambda(M(\theta'))$ of the certificate $M(\theta')$. Second, the above problem \eqref{eq:prin_prob} bears a similarity to the classic work of \citet{mussa1978}. The key difference is that the platform is choosing \emph{both} the quality ($\Lambda$) and quantity ($V_g$, $V_b$) of the product, and that these two are related.

	\subsection{Preliminary Analysis and Simplification of the Platform's Problem}
	
	We simplify the platform's problem with the following observation.
	
	\begin{restatable}{lemma}{simplifcationlemma}
		\label{mutheta}
		For any incentive compatible and individually rational mechanism $(V_g,V_b,M,P)$ there exists another incentive compatible and individually rational mechanism $(V_g,\tilde{V}_b,\tilde{M},P)$ such that, for all $\theta\in \Theta$, $\tilde{M}(\theta)=\theta$ and, both the platform and good providers of every value receive the same payoff as from $(V_g,V_b,M,P)$.
	\end{restatable}
	
	
	In words, this lemma states that it is without loss to assign a separate certificate to each value $\theta$. This is because different certificates can have the same quality. So we can take any mechanism in which different values are assigned to the same certificate and construct a new mechanism in which all values have distinct certificates but we reassign the bad providers' views in a way that the quality is constant across certificates. The intuition stems from the linearity of the platform's payoff in $V_b$; the platform can reallocate the bad provider views $V_b$ freely across different messages with identical qualities such that the ratio of good to bad content is equal to the average level, value-by-value.
	
	This allows us to drop the certification decision $M$ and the function $V_b$ determining the bad provider views and rewrite the platform's problem with it choosing the quality $\Lambda:\Theta\to [0,1]$ of the certificate for each good provider value $\theta$. This is a consequence of Lemma \ref{mutheta} and the fact that the bad provider views $V_b$ are pinned down by the equation $\Lambda(\theta)=V_g(\theta)/[V_g(\theta)+V_b(\theta)]$. The platform thus solves\footnote{In what follows, note that, in the integrand, a fraction whose numerator and denominator are both zero takes the value zero. We also note that, as written, the problem permits us to choose $\Lambda(\theta)>0$ and $V_g(\theta)=0$ (which technically violates the definition of $\Lambda$). We do not explicitly prevent this by imposing an additional constraint as, when $V_g(\theta)=0$, the objective function and the constraints take the same values for any $\Lambda(\theta)\in[0,1]$. This mild abuse allows us to state results more concisely without changing their economic content.}
	\begin{align*}
		\begin{split}
			& \max_{V_g,\Lambda,P} \int_{\Theta} \left[ P(\theta)+ A(\Lambda(\theta))V_g(\theta)\frac{1-\Lambda(\theta)}{\Lambda(\theta)} -\gamma\frac{V_g(\theta)}{\Lambda(\theta)} - c(V_g(\theta)) \right]f(\theta)d\theta, \\
			& \text{subject to}\\
			& \theta A(\Lambda(\theta))V_g(\theta) - P(\theta) \geq \max\{\theta A(\Lambda(\theta'))V_g(\theta') - P(\theta'),0\}\qquad \text{ for all }\theta,\theta'\in \Theta.
		\end{split}
	\end{align*}
	
	If we interpret $A(\Lambda(\theta))V_g(\theta)$ as the ``allocation'' when value $\theta$ is reported, the incentive compatibility constraint is essentially identical to the standard incentive compatibility constraint of \citet{mussa1978}. Thus, we can use the standard characterization of incentive compatibility to eliminate the price function $P$ from the platform's problem and restate it as
	\begin{align}\label{eq:prin_prob_simple}
		\begin{split}
			& \max_{V_g,\Lambda} \int_{\Theta} \left[ \left(\phi(\theta)+\frac{1-\Lambda(\theta)}{\Lambda(\theta)}\right) A(\Lambda(\theta))V_g(\theta) - \gamma\frac{V_g(\theta)}{\Lambda(\theta)} - c(V_g(\theta)) \right]f(\theta)d\theta, \\
			& \text{subject to}\\
			& A(\Lambda(\theta))V_g(\theta)\geq A(\Lambda(\theta'))V_g(\theta')\;\; \text{ for } \theta\geq \theta',\; \theta,\theta'\in \Theta.
		\end{split}
	\end{align}
	In the above objective function $$\phi(\theta):=\theta-\frac{1-F(\theta)}{f(\theta)}$$ is the standard \textit{virtual value}. The constraint captures the fact that incentive compatibility requires the allocation to be nondecreasing. Note that we also eliminated the individual rationality constraint in the standard way by observing that it is optimal for the platform to provide a utility of zero to a good provider of value $\theta=0$. We refer to a solution of the platform's problem \eqref{eq:prin_prob_simple} as an \textit{optimal mechanism}.
	
	The following lemma summarizes the simplification of the platform's problem. It states that a pointwise solution that satisfies the monotonicity required for incentive compatibility is an optimal mechanism.
	\begin{lemma}\label{lem:pointwise}
		Suppose there is a mechanism $(V^p_g,\Lambda^p)$ such that
		$$(V^p_g(\theta),\Lambda^p(\theta))\in \argmax_{v_g,\lambda} \left[ \left(\phi(\theta)+\frac{1-\lambda}{\lambda}\right) A(\lambda)v_g - \gamma\frac{v_g}{\lambda}  - c(v_g)\right]$$
		pointwise maximizes the platform's objective for all $\theta\in \Theta$ and that $A(\Lambda^p(\theta))V^p_g(\theta)$ is nondecreasing in $\theta$. Then $(V^p_g,\Lambda^p)$ is an optimal mechanism; that is, it is a solution to the platform's problem \eqref{eq:prin_prob_simple}.		
	\end{lemma}
	
	In what follows, we assume the distribution $F$ is such that the virtual value $\phi$ is strictly increasing. This is a standard technical assumption in mechanism design. This assumption simplifies the derivation of the optimal mechanism; as we will show, it implies that there is a pointwise optimum $(V^p_g,\Lambda^p)$ as described in Lemma \ref{lem:pointwise} such that $A(\Lambda^p(\theta))V^p_g(\theta)$ is indeed nondecreasing. This assumption can be dispensed with by employing standard ironing techniques but we choose not to do so because we view the additional technicality (required for this generality) complicates the presentation of our main economic insights.
	
	A useful way to view the pointwise problem in Lemma \ref{lem:pointwise} is to separate the decision of certificate quality from the decision of the quantity of views to allocate to good providers. Let
	$$R(\hat{\phi},\lambda):=\left(\hat{\phi}+\frac{1-\lambda}{\lambda}\right) A(\lambda) - \frac{\gamma}{\lambda}.$$
	Then, $\Lambda^p(\theta)=\argmax_{\lambda} \{R(\phi(\theta),\lambda)\}$ and $V^p_g(\theta)=\argmax_{v_g}\{R(\phi(\theta),\Lambda^p(\theta))v_g-c(v_g)\}$. The latter problem, of choosing $v_g$, mimics \citet{mussa1978}, but where the maximized $R$ stands in for the effective virtual value (or return) generated by $v_g$. Notice that when $\lambda=1$, $R(\phi(\theta),1)=\phi(\theta)-\gamma$, that is, the function $R$ captures the true virtual value net of the cost $\gamma$ of an untargeted view. The platform will deviate from perfect certification because this can lead to a higher value of $R(\phi(\theta),\lambda)$. The quantity of views directed at content the of good providers is simply derived by equating the marginal cost of targeting, $c'$, with the return $R(\phi(\theta),\Lambda^p(\theta))$ if it is positive, and is set to zero if $R(\phi(\theta),\Lambda^p(\theta))$ is negative.
	
	\section{Benchmarks}\label{sec:benchmarks}
	
	In this section, we derive three benchmarks that provide context for the properties of the optimal mechanism. We first solve a planning problem in which the goal is to maximize consumer engagement net of platform costs. We then derive the properties of the optimal mechanism when the platform is restricted to using a constant quality for all $\Theta$. This includes the case of perfect quality---not mixing content from good and bad providers---as desired by the European Commission. This contrasts the maximization of consumer engagement and platform profits. We then derive the properties of the optimal mechanism when the platform is restricted to offering only two certificates (so the quality function can take at most two values). This allows us to transparently demonstrate the value to the platform of selling multiple certificates, which is typically a feature of the solution to platform's problem described in expression (\ref{eq:prin_prob}).
	
	\subsection{Engagement Maximizing Planner}\label{sec:planner}
	
	We consider a planner who wants to maximize the level of engagement on the platform net of platform costs.\footnote{Under the particular class of attention functions $A(\lambda)=\lambda^{\alpha}$, user welfare is described by engagement (up to a constant of proportionality). See Appendix \ref{sec:appendixwelfare} for details.} That is, the planner solves
	\begin{equation*}
		\max_{V_g,\Lambda} \left\{\int_{\Theta} \left[ A(\Lambda(\theta))V_g(\theta) -\gamma\frac{V_g(\theta)}{\Lambda(\theta)}  - c(V_g(\theta))\right]f(\theta)d\theta.\right\}
	\end{equation*}
	The first term in the objective function above is the engagement $A(\Lambda(\theta))V_g(\theta)$ of users with good providers (not the content provider's utility which incorporates the value of that engagement and writes as $\theta A(\Lambda(\theta))V_g(\theta)$). The remaining terms are the costs: both targeting costs for views to good providers, and untargeted views to bad providers (recalling that $V_b(\theta)=V_g(\theta)\frac{1-\Lambda(\theta)}{\Lambda(\theta)}$). 
	
	Note that the planner's objective is not the same as the sum of the utilities of the consumers of content and the profit of the platform. To determine the former, we would need to take a stand on the consumer utility from engagement generating content (which we normalized to 1); this combined with the attention function $A$ pins down the distribution of the time costs of content reading $q$. Since consumers do not pay for content, there is no natural way to relate consumer preferences and firm profits (as there is in standard mechanism design problems). We therefore focus on features that are general to the relative weighting of the costs and benefits for the planner. Indeed, the social surplus should also incorporate the payoff of the content providers but we do not include this as content provider utility is never brought up in the policy debate around certification for sale. 
	
	The solution $(V^e_g,\Lambda^e)$ to the problem of maximizing engagement above is immediate from pointwise optimization (since there are no constraints) and is given by
	\begin{align*}
		& \Lambda^e(\theta) = 1, \\
		& V^{e}_g(\theta)=c'^{-1}(1-\gamma)
	\end{align*}
	for all $\theta\in\Theta$.\footnote{Recall that we have normalized $A(1)=1$.}
	
	We flag two intuitive properties of this solution. The first is that no views are directed to bad providers since directing content to them is costly and does not generate engagement. In other words, certification for all $\theta\in\Theta$ is perfect with all certificates having quality one. Second, since the planner only wants to maximize engagement, the same number of views are directed to good providers regardless of their value $\theta$. Such egalitarian traffic will not arise from a profit maximizing platform, since the platform will direct more traffic to providers with higher willingness to pay for those views. Note that these two qualitative properties would be unchanged if we alter the planner's objective to assign different weights to the total engagement and platform costs.
	
	\subsection{One certificate}\label{subsec:simple}
	
	In this section, we study a benchmark in which the platform assigns the same message to all types $\theta\in\Theta$ or, in terms of the simplified platform problem in expression \eqref{eq:prin_prob_simple}, the function $\Lambda$ is restricted to be a constant function that takes some value $\lambda\in [0,1]$. 
	
	We do this for several reasons. First, this corresponds to a realistic form of content moderation whereby platforms do not distinguish between different kinds of content (that is, all content is presented in the same way) but still choose how many views to allocate to different providers. The platform may choose to ban bad provider traffic (perfect certification) or not (imperfect certification). Second, the analysis crisply illustrates how imperfect certification can raise platform profits by expanding the subset of types $\Theta$ that the platform profitably serves. In this sense, imperfect certification allows for greater content diversity-—a central theme of our analysis and one that features in the optimal mechanism. Finally, it allows us to characterize the solution if the platform is forced to certify perfectly; that is, no content from bad providers is assigned the same certificate as content from good providers.
	
	We first derive the optimal number of views $V^g$ for an arbitrary quality $\lambda\in[0,1]$. We then vary the quality $\lambda$ to demonstrate the cross-subsidization effect of bad content on good content, and to consider a hypothetical policy that limits selling of certification without restricting steering. We view this as the natural benchmark for how the European Commission claims platforms should operate: certification should not be for sale, and certificates should be a clear statement of quality (as, for instance, they claim should be the case on X).\footnote{ \url{https://ec.europa.eu/commission/presscorner/detail/en/ip_24_3761}} They have not taken or suggested any action against platforms that sell traffic in various ways, however.
	
	Fixing a $\lambda$, the platform's problem \eqref{eq:prin_prob_simple} is
	\begin{align}\label{eq:SingleProblem}
		\begin{split}
			\Pi^s(\lambda):=&\max_{V_g}\left\{\int\left(\left(\phi(\theta)+\frac{1-\lambda}{\lambda}\right)A(\lambda)V_{g}(\theta)-\gamma\frac{V_{g}(\theta)}{\lambda}-c(V_{g}(\theta))\right)f(\theta)d\theta\right\}\\
			& \text{subject to}\\
			& V_g(\theta)\geq V_g(\theta')\;\; \text{ for } \theta\geq \theta',\; \theta,\theta'\in \Theta.
		\end{split}
	\end{align}
	
	It is immediate here that the pointwise optimum 
	$$V^{s,\lambda}_{g}(\theta)=c'^{-1}\left(\max\left\{\left(\phi(\theta)+\frac{1-\lambda}{\lambda}\right)A(\lambda)-\frac{\gamma}{\lambda},0\right\}\right)=c'^{-1}(\max\{R(\phi(\theta),\lambda),0\})$$
	satisfies the required monotonicity constraint because we assumed $\phi$ is increasing. For $\theta\in\Theta$ such that $V^{s,\lambda}_g(\theta)$ is not zero, the value is derived from the first-order condition
	\begin{equation}\label{eq:FOCVsimple}
		c'(V^{s,\lambda}_{g}(\theta))=\left(\phi(\theta)+\frac{1-\lambda}{\lambda}\right)A(\lambda)-\frac{\gamma}{\lambda}=R(\phi(\theta),\lambda).
	\end{equation}
	
	A useful special case is perfect certification, $\lambda=1$, in which only good providers are assigned views. This sort of perfect certification would avoid a policy maker's complaint that messages are ``deceiving.'' So, suppose that perfect certification were enforced, but steering was still for sale.\footnote{This would seem to be consistent with the European Commission's concern and possible action regarding X, for example.} Then, the optimal number of views under perfect certification is $V^{s,1}_{g}(\theta)=c'^{-1}\left(\max\{\phi(\theta)-\gamma,0\}\right).$
	
	This solution also provides a natural connection to classic price discrimination: it exactly mirrors \citet{mussa1978}. Since there are only good content providers, the benefit is the virtual value, and the cost is the sum of the opportunity cost $\gamma$ and the targeting cost $c$. In this solution, the marginal cost of assigning an additional view to a content provider of value $\theta$ is equal to the benefit which is precisely the virtual valuation $\phi$ (that ensures the appropriate information rents accrue to the good providers). Compared to engagement maximization, where all content gets the same number of views, the monopoly platform creates an asymmetry in the views allocated across $\theta$ via the virtual valuation $\phi$. Those good providers with higher $\theta$ who value being seen more reveal their higher valuation by paying more to enjoy a higher number of engaged views. 
	
	Following the first-order condition \eqref{eq:FOCVsimple}, imperfect certification has two effects. First, it lowers attention and therefore reduces the amount that can be charged to good types; this reduction is proportional to the virtual value. On the other hand, it generates, for every good view, some bad traffic that can be monetized. This effect is constant and raises the payoff from views symmetrically for all virtual values. Therefore imperfect certification increases the value of views relatively more for low virtual values. As a result, imperfect certification makes traffic more egalitarian.
	
	The following result provides a sufficient condition for the profit maximizing simple certification to be imperfect.
	
	\begin{restatable}{proposition}{singleone}
		\label{prop:simple}
		Suppose $\overline{\theta} A'(1) < 1 - \gamma $. Then, every optimal single certificate $$\lambda^{s^*}\in \argmax_{\lambda}\left\{\Pi^{s}(\lambda)\right\}$$ satisfies $\lambda^{s^*} < 1$. That is, optimal single certification is imperfect.
	\end{restatable}
	If $A'(1)$ is not too big, then the gains from imperfect certification outweigh the costs of lost attention and engagement. The net return to bad provider traffic near perfect certification, $1-\gamma$, provides the sufficient bound on $A'(1)$.
	
	Imperfect certification, when it is desirable for the platform, raises content diversity. Formally, we define the \textit{content diversity} of a mechanism $(V_g,\Lambda)$ as the set of good provider values $\{\theta\in\Theta|V_g(\theta)>0\}$ that are allocated views. Thus, a mechanism $(V'_g,\Lambda')$ has greater content diversity than another mechanism $(V_g,\Lambda)$ if a larger set of values are served by the former or $\{\theta\in\Theta|V'_g(\theta)>0\}\supset \{\theta\in\Theta|V_g(\theta)>0\}$. The following result shows that either the platform prefers perfect certification over a fixed level of imperfect certification, or the imperfect certification increases content diversity.
	
	\begin{restatable}{proposition}{singletwo}
		For any $0<\lambda <1$, either $\Pi^s(\lambda)<\Pi^s(1)$, or content diversity is higher under $\lambda$ then under perfect certification.
	\end{restatable}
	This previews our result which will extend to the full mechanism below: enforcing perfect certification, when it is not optimal for the platform, lowers content diversity.
	
	\subsubsection*{Linear Attention}
	Throughout the paper, we will use the example $A(\lambda)=\lambda$ of a linear attention function (which corresponds to a uniform distribution of the cost of reading content) to illustrate the underlying forces driving our results. In particular, here, we use it to demonstrate how greater content diversity can arise with imperfect certification.
	
	For a fixed level of certification $\lambda$, the corresponding optimal $V^{s,\lambda}_g$ is given by $$V^{s,\lambda}_{g}(\theta)=c^{-1}\left(\max\left\{\phi(\theta)\lambda+1-\lambda-\frac{\gamma}{\lambda},0\right\}\right).$$
	Perfect certification implies $\lambda=1$ and therefore, $V^{s,1}_{g}(\theta)=c^{-1}(\max\{\phi(\theta)-\gamma,0\})$.
	
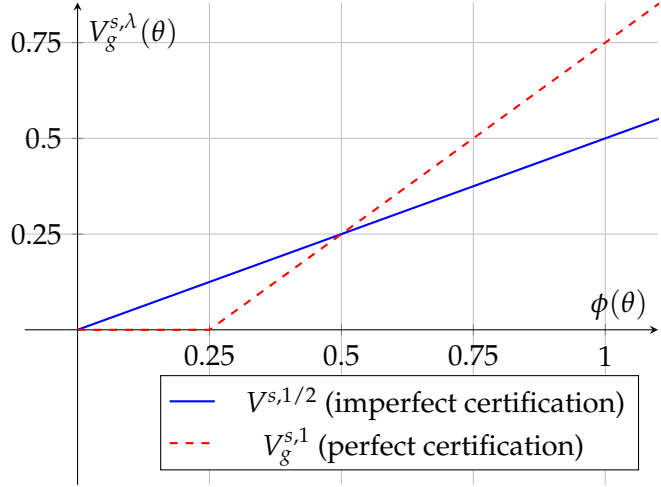
\begin{figure}[h]
		\begin{center}
			\begin{tikzpicture}     \begin{axis}[         width=10cm,         height=8cm,         xlabel={$\phi(\theta)$},         ylabel={$V^{s,\lambda}_g(\theta)$},         xmin=0, xmax=1,         ymin=-0.3, ymax=0.75,         xtick={0,0.25,0.5,0.75,1},         ytick={0,0.25,0.5,0.75},         grid=both,         major grid style={line width=.2pt,draw=gray!50},         axis lines=middle,         enlargelimits,         legend pos=south east     ]     \addplot[         domain=0:2,          samples=400,          thick,         blue,     ] {max((x+1)/2 - 0.5, 0)};  \addlegendentry{$\;\;\;V^{s,1/2}$ (imperfect certification)} ;\addplot[dashed,         domain=0:2,          samples=400,          thick,         red,     ] {max(x - 0.25, 0)};    \addlegendentry{$V^{s,1}_g$ (perfect certification)}      \end{axis} \end{tikzpicture}
			\par\end{center}%
		\caption{Good provider views with simple certification. Parameters: $A(\lambda)=\lambda$, $\gamma=1/4$ and $c(v_g)=v_g^{2}/2$.}
		\label{fig:single}
	\end{figure}

	Figure \ref{fig:single} shows the number of views a good content provider enjoys (on the y-axis) as a function of their virtual value (and therefore, indirectly of their type) and compares perfect certification (depicted by the dashed red line) and imperfect certification with a single certificate of quality $\lambda=\frac{1}{2}$ (depicted by the blue line).
	
	The figure highlights the two senses discussed above in which imperfect certification results in an allocation of views that is closer to the engagement maximizing benchmark (recall that the engagement maximizing benchmark has identical traffic for all $\theta$). First, note that there is greater content diversity under imperfect certification; specifically, more lower value good providers are served with imperfect certification. This comes from good providers being served at a loss. Good providers with values that satisfy $\phi(\theta)A(\lambda)<\gamma$ are unwilling to make payments high enough to cover their average cost; they are served to keep attention high enough to make it worth serving bad types. Second, traffic is more egalitarian in that the number of views are less sensitive to the virtual value (and therefore the value) with imperfect certification (observe that the red line corresponding to perfect certification is steeper). Of course, these notions of improved content diversity do not translate directly into higher total engagement or welfare. The aggregate comparison is not immediately obvious from the figure since, in shifting from perfect to imperfect certification, some (higher) values receive fewer and other (lower) values receive more views. Finally, all good providers receive less attention conditional on being viewed. Moreover, in this figure, we have not taken a stance on the distribution $F$ of types. We take up welfare in more detail below. However it is immediate that if the mass of $F$ is concentrated on lower values then engagement must be higher with imperfect certification.
	
	\subsection{Two certificates}\label{sec:two_certificates}
	
	Before moving on to our analysis of the general problem, we build some intuition by considering the case of two certificates. This case may also be of independent interest in that it mirrors recent developments in many social media platforms. Specifically, until relatively recently, Instagram and Twitter (now X) had users who were either verified (and their accounts were marked with a check sign) or not (their accounts were unmarked). While they initially did not charge for those providers who obtained a verified status, two certificates is a natural benchmark to study due to this historical precedent. Indeed, when Twitter first started charging for the provision of verified status, they only (in our language) offered two certificates. This section shows how this practice can improve profits for the platform.
	
	The previous subsection showed that fixed imperfect (relative to perfect) certification can lead to higher profits and greater content diversity. The case with two certificates further illustrates how multiple levels of certification can raise the profits for the platform. By using two certificates, the platform can profitably serve low-value good providers, by cross subsidizing them with bad providers, while simultaneously not sacrificing engagement for the views of high value good providers.\footnote{In using certification to soften the incentive constraints for higher types, there is some similarity to \citet{deneckere1996damaged}. Of course, in our environment the platform earns revenue (from bad providers) in ``damaging" the good rather than incurring costs. More substantively, in \citet{deneckere1996damaged} consumers are constrained to unit demand, whereas we vary both the number of views and the quality of the certificate. This results in different economic effects.} Furthermore, the two certificate benchmark allows us to demonstrate how the two instruments---certification and steering---interact. We provide conditions under which the optimal policy has two distinct levels of certification, and many levels of steering; simple imperfect certification may be profitable, but is not optimal.
	
	As for the case of a single certificate above, we begin by supposing that the quality associated with the two certificates is exogenously given by $\umu,\omu\in[0,1]$ with $\umu \leq \omu$. We write the two certificate benchmark problem for the platform as
	\begin{equation}\label{eq:two_certificates}
		\begin{split}
			\Pi^{bin}(\umu,\omu,\hat{\theta}):= \max_{V_g} &\left\{ \int_0^{\hat{\theta}} \left[ \left(\phi(\theta)+\frac{1-\umu}{\umu}\right) A(\umu)V_g(\theta) -\gamma\frac{V_g(\theta)}{\umu} - c(V_g(\theta)) \right]f(\theta)d\theta \right. \\
			&\left.  +\int_{\hat{\theta}}^{\overline{\theta}} \left[ \left(\phi(\theta)+\frac{1-\omu}{\omu}\right) A(\omu)V_g(\theta) -\gamma\frac{V_g(\theta)}{\omu} - c(V_g(\theta)) \right]f(\theta)d\theta \right\},
		\end{split}
	\end{equation}
	where $\hat{\theta}\in\Theta$. In the above problem, all values above and below $\hat{\theta}$ are assigned the higher $\omu$ and lower $\umu$ quality certificates respectively.
	
	The optimal two certificate mechanism can be derived by choosing the appropriate $\umu$, $\omu$ and $\hat{\theta}$ to maximize $\Pi^{bin}(\umu,\omu,\hat{\theta})$.\footnote{We characterize the two certificate optimal mechanism in Appendix \ref{sec:appendixtwocert}; we choose not present the result here for brevity. We show that the optimal two certificate mechanism  takes the cutoff form assumed in the above problem \eqref{eq:two_certificates}.} In addition to higher value good providers receiving the higher quality certificate, they also receive more targeted views. Thus, as in the single certificate optimum, the platform uses the quantity of views to price discriminate but now also uses the certificate quality to separate higher from lower value good providers. The platform assigns higher-quality certificates to high-value good providers so as not to dilute earnings from these types, while assigning lower-quality certificates to use low-value good providers to use them as a means of earning revenue from the bad providers. 
	
	Instead of presenting the two certificate optimum, we derive a sufficient condition under which the platform gets strictly higher profits from using two certificates (as opposed to a single certificate $\umu=\omu$). 
	\begin{restatable}{proposition}{twocertificates}
		\label{prop:two_certificates}
		Suppose $\overline{\theta}A'(1)<1-\gamma$. Then, every two certificate optimum $$(\umu^*,\omu^*,\hat{\theta}^{bin^*})\in\argmax_{\umu,\omu,\hat{\theta}} \;\left\{\Pi^{bin}(\umu,\omu,\hat{\theta})\right\}$$ satisfies $\umu^* < \omu^*$.
	\end{restatable}
	
	This is the same sufficient condition under which the platform chooses imperfect certification when restricted to a single certificate (Proposition \ref{prop:simple}), so some form of imperfect certification must be optimal. The reason for using multiple certificates is that, in the platform's problem \eqref{eq:two_certificates}, the profit depends on the interaction of the good provider's value with the quality.
	
	We next turn to the fully optimal mechanism in which any number of messages can be used. The forces outlined in both the single and two certificate cases come to the fore: the platform uses different quantities of views to price discriminate coupled with the assignment of distinct quality certificates to effectively extract revenue from bad providers.
	
	\section{Certification and steering for sale}
	
	\subsection{The optimal mechanism}
	
	We now characterize and analyze the optimal mechanism. The characterization builds on the intuition in Section \ref{sec:benchmarks}. Relative to the planner's problem, the profit-seeking platform may use imperfect certificates as a means of raising revenue from bad providers and can more profitably price discriminate across good content providers by offering different quantities of targeted views and distinct quality certificates. The following proposition shows that the pointwise optimum satisfies the required conditions to ensure incentive compatibility and both the views $V_g$ and the quality $\Lambda$ are nondecreasing in $\theta$.\footnote{When $V^*_g(\theta)=0$, the platform makes no profits from value $\theta$ so any $\lambda$ is optimal. For expositional simplicity, we pick the $\lambda$ that maximizes effective virtual value $R(\phi(\theta),\lambda)$. Further, whenever there are multiple values of $\lambda$ that maximize $R(\phi(\theta),\lambda)$, we pick the largest solution. This choice is convenient to present results and does not sacrifice any qualitative insight.}
	
	\begin{restatable}{proposition}{optmech}
		\label{mylemma}
		\label{prop:opt_mech}
		There is an optimal mechanism $(V^*_g,\Lambda^*)$ solving the platform's problem \eqref{eq:prin_prob_simple} where both $V^*_g$, $\Lambda^*$ are nondecreasing and satisfy
		\begin{align*}
			& \Lambda^*(\theta)=\max\left\{\tilde{\lambda}\;\;\bigg|\;\; \tilde{\lambda}\in \argmax_{\lambda\in [0,1]} \left\{\left(\phi(\theta)+\frac{1-\lambda}{\lambda}\right)A(\lambda)-\frac{\gamma}{\lambda}\right\}\right\} \;\;  \;\;  \\
			& V^*_g(\theta)=c'^{-1}\left(\max\left\{\left[\phi(\theta)+\frac{1-\Lambda^*(\theta)}{\Lambda^*(\theta)}\right]A(\Lambda^*(\theta))-\frac{\gamma}{\Lambda^*(\theta)},0\right\}\right)
		\end{align*}
		for all $\theta\in \Theta$.
	\end{restatable}
	
	
	Recall that incentive compatibility is satisfied if the product $A(\Lambda^*(\cdot))V^*_g(\cdot)$ is nondecreasing; thus, one qualitative contribution of the above result is to show that both $\Lambda^*(\cdot)$ and $V^*_g(\cdot)$ are each individually nondecreasing. This implies that good providers with higher values both receive more views and their content is pooled with less bad content. Certification can be perfect (that is, $\Lambda^*(\theta)=1$) for sufficiently high values $\theta$. Recall that one can interpret this as first choosing $\lambda$ to maximize the effective virtual value
	$$\Lambda^*(\theta)=\max\left\{\tilde{\lambda}\;\;\bigg|\;\; \tilde{\lambda}\in \argmax_{\lambda\in [0,1]} \{R(\phi(\theta),\lambda)\} \right\}$$
	and then choosing the views in the ``standard" way as $$V^*_g(\theta)=c'^{-1}(\max\{R(\phi(\theta),\Lambda^*(\theta)),0\}).$$
	
	The forces in Section \ref{subsec:simple} that pushed the optimal single certificate mechanism with imperfect (relative to perfect) certification qualitatively closer to planner's optimal solution also apply in the optimal mechanism of Proposition \ref{prop:opt_mech}. Serving low-value good providers allows the platform to earn more revenue from bad providers. This, in turn, flattens out the relationship between targeted views and good provider valuations leading to more egalitarian content provision relative to perfect certification.
	
	The optimal mechanism of Proposition \ref{prop:opt_mech} features greater content diversity than either the single or two certificate benchmarks. To see this, it is instructive to compare the optimal (unrestricted) mechanism to the mechanism that maximizes profits when the platform is restricted to offer only two certificates (derived in Proposition \ref{prop:twocert} in the appendix). For any pair of binary qualities $0< \umu<\omu\leq 1$, it must be the case that
	$$\max_{\lambda\in [0,1]} \left\{\left(\phi(\theta)+\frac{1-\lambda}{\lambda}\right)A(\lambda)-\frac{\gamma}{\lambda}\right\}\geq \max_{\lambda\in \{\umu,\omu\}} \left\{\left(\phi(\theta)+\frac{1-\lambda}{\lambda}\right)A(\lambda)-\frac{\gamma}{\lambda}\right\}$$
	and consequently, the set of values $\{\theta\in \Theta\;|\; V^*_g(\theta)>0\}\supseteq \{\theta\in \Theta\;|\; V^{bin}_g(\theta)>0\}$ which receive any views at all is a larger set in the optimal mechanism relative to the binary benchmark. This greater content diversity is a result of the optimal mechanism directing traffic towards low-values  $\theta$ that are unserved under binary certificates. These are values $0\leq \theta\leq \underline{\theta}$ for which
	$$\max_{\lambda\in [0,1]} \left\{\left(\phi(\theta)+\frac{1-\lambda}{\lambda}\right)A(\lambda)-\frac{\gamma}{\lambda}\right\}>0$$ where $\underline{\theta}$ satisfies
	$$\left(\phi(\underline{\theta})+\frac{1-\umu}{\umu}\right)A(\umu)-\frac{\gamma}{\umu}=0.$$
	
	\subsubsection*{Linear Attention}
	
	Returning to the case where the attention function $A(\lambda)=\lambda$ is linear, Proposition \ref{prop:opt_mech} implies that the mechanism
	\begin{equation}\label{eq:linear_opt}
		\begin{split}
			& \Lambda^*(\theta)=\left\{\begin{array}{cl}
				\sqrt{\frac{\gamma}{1-\phi(\theta)}} & \text{ if  } \phi(\theta)\leq1-\gamma,\\
				1 & \text{ if  } \phi(\theta)>1-\gamma,
			\end{array}\right.\\
			& V^*_{g}(\theta)=\left\{\begin{array}{cl}
				c'^{-1}\left(\phi(\theta)\Lambda^*(\theta)+1-\Lambda^*(\theta)-\frac{\gamma}{\Lambda^*(\theta)}\right) &\text{ if  } \phi(\theta)\geq 1-\frac{1}{4\gamma}\; \text{ and }\\
				0 &\text{ if  } \phi(\theta)<1-\frac{1}{4\gamma}
			\end{array}\right.
		\end{split}
	\end{equation}
	is optimal.
	
	Now observe that $1-\frac{1}{4\gamma}\geq 1-\gamma$ when $\gamma\geq 1/2$. Consequently, when $\gamma\geq$ 1/2, the optimal mechanism has the property that $\Lambda^*(\theta)=1$ for all $\theta\in\Theta$ such that $V^*_{g}(\theta)>0$. That is, perfect certification is optimal when it is sufficiently costly to supply untargeted views. We therefore focus on the case of $\gamma<1/2$ in which the optimal mechanism features a variety of levels of certification.
	
\begin{figure}[h]
		\begin{subfigure}[t]{\textwidth}
			\centering
			\begin{tikzpicture}     \begin{axis}[         width=10cm,         height=8cm,         xlabel={$\phi(\theta)$},         ylabel={$\Lambda(\theta)$},         xmin=0, xmax=1,         ymin=-0.3, ymax=1,         xtick={0,0.25,0.5,0.75,1},         ytick={0,0.25,0.5,0.75,1},         grid=both,         major grid style={line width=.2pt,draw=gray!50},         axis lines=middle,         enlargelimits,         legend pos=south east     ]     \addplot[         domain=0:2,          samples=400,          thick,         blue,     ] {min(0.5*(1-x)^-0.5, 1)};    \addlegendentry{$\Lambda^*$, the optimal certificate quality}  ;     \end{axis} \end{tikzpicture}
			\caption{Certificate quality}\label{fig:opt_mech_linear_qualitypanel}
		\end{subfigure}
		\begin{subfigure}[t]{\textwidth}
			\centering
			\begin{tikzpicture}     \begin{axis}[         width=10cm,         height=8cm,         xlabel={$\phi(\theta)$},         ylabel={$V_g(\theta)$},         xmin=0, xmax=1,         ymin=-0.3, ymax=0.75,         xtick={0,0.25,0.5,0.75,1},         ytick={0,0.25,0.5,0.75},         grid=both,         major grid style={line width=.2pt,draw=gray!50},         axis lines=middle,         enlargelimits,         legend pos=south east     ]   ;  \addplot[         domain=0:0.75,          samples=400,          thick,         blue,     ] {1 - sqrt(1 - x))}; \addlegendentry{$V^*_g$, the optimal targeted views}  ;\addplot[dashed,         domain=0:2,          samples=400,          thick,         red,     ] {max(x - 0.25, 0)};    \addlegendentry{$V^{s,1}_g$ (perfect certification)} ;\addplot[         domain=0.75:2,          samples=400,          thick,         blue,     ]  {max(x - 0.25, 0)};  \end{axis}    \end{tikzpicture}
			\caption{Content Diversity}\label{fig:opt_mech_linear_Vpanel}
		\end{subfigure}
		\caption{The optimal mechanism. Parameters:  $A(\lambda)=\lambda,\gamma=1/4$, $c(v_g)=v^2_g/2$.}\label{fig:opt_mech_linear}
	\end{figure}
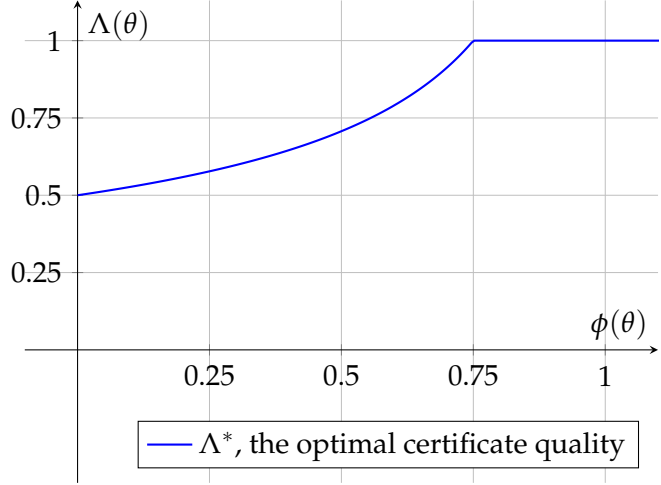
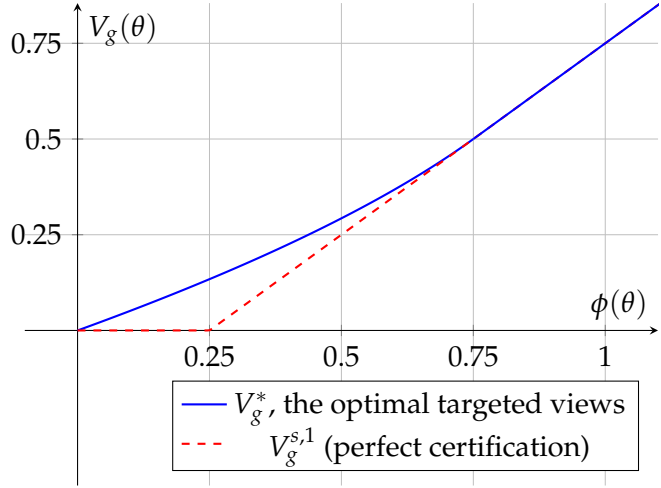

	Figure \ref{fig:opt_mech_linear} illustrates such a case. Each panel has the good provider's virtual value $\phi$ on the x-axis. Panel \ref{fig:opt_mech_linear_qualitypanel} plots the certificate quality $\Lambda^*$ and Panel \ref{fig:opt_mech_linear_Vpanel} plots the number of good provider views $V^*_{g}$ in the optimal mechanism. Sufficiently high values are all assigned perfect certificates but the platform still price discriminates through the number of views it provides. The platform employs both instruments---imperfect certification and the quantity of views---for the lower values. In contrast to the case with only two certificates, the sale of lower quality certificates to lower values entails a gradual, continuous degradation in certificate quality (rather than a discrete fall).
	
	As a result of using different certificates to price discriminate, content diversity increases relative to perfect certification. In addition, values $\theta$ that are assigned imperfect certificates ($\Lambda^*(\theta)<1$) receive more views than they would under perfect certification. As in the case with a single imperfect certificate, good providers with values $\theta$ such that $\phi(\theta)A(\Lambda(\theta))<\gamma$ are being served at a loss, in terms of direct revenue from the prices charged, in order to generate revenue from bad providers.\footnote{Notice that the fact that all good providers served in this example have positive virtual values is an artifact of the specific chosen parameters. For a lower $\gamma$, some values $\theta$ with negative virtual values $\phi(\theta)<0$ are also served ($V^*_g(\theta)>0$) in the optimal mechanism.}
	
	\subsection{Comparative Statics}
	The solution described in Proposition \ref{prop:opt_mech} allows us to conduct several comparative statics exercises. They help to explain the circumstances under which imperfect certification is optimal. Lower costs $\gamma$ for untargeted views or a more concave attention function $A$ make optimal certification more imperfect but a proportionate lower cost of targeting $c$ has no impact.
	
	\paragraph{\emph{Costs of untargeted views}}
	We first examine how the quality of the certificates in the optimal mechanism $(V^*_g,\Lambda^*)$ are affected by the cost $\gamma$. Recall, that a natural interpretation for $\gamma$, which is the opportunity cost of providing an untargeted view, is lost ad revenue. Thus, the next result shows that when ad revenue falls, good content providers enjoy worse certificates.
	
	\begin{restatable}{proposition}{compstaticsgamma}
		\label{prop:comp_statics_gamma}
		The quality $\Lambda^*(\theta)$ in the optimal mechanism $(V^*_g,\Lambda^*)$ is nondecreasing in $\gamma$ for all $\theta\in\Theta$.
	\end{restatable}
	
	
	For intuition, consider a good provider with value $\theta\in \Theta$ that is assigned to an imperfect certificate $\Lambda^*(\theta)\in(0,1)$. Since, $\Lambda^*(\theta)$ maximizes $\left(\phi(\theta)+\frac{1-\lambda}{\lambda}\right)A(\lambda)-\frac{\gamma}{\lambda}$ (see Proposition \ref{prop:opt_mech}), it satisfies the first-order condition
	\begin{equation}{\label{eq:FOCmu}}
		A(\Lambda^*(\theta))-\Lambda^*(\theta)^{2}\left(\phi(\theta)+\frac{1-\Lambda^*(\theta)}{\Lambda^*(\theta)}\right)A'(\Lambda^*(\theta))=\gamma.
	\end{equation}
	This equation captures the tradeoffs of including additional bad content with a given certificate. The right hand side is the cost of the additional bad view. If $\gamma$ is lowered, more bad provider views can be absorbed before the left side balances the right. The benefits of the bad traffic on the left contains two terms. The first is the amount $A(\Lambda^*(\theta))$ that a bad provider pays for this view. But including more bad content reduces the quality of the certificate and therefore the attention; this, in turn, reduces the amount that both good and bad providers are willing to pay, which is captured by the second term. 
	
	On the margin, additional bad views reduce the quality by $\frac{\Lambda^*(\theta)^{2}}{V^*_g(\theta)}$ and therefore the attention by $\frac{\Lambda^*(\theta)^{2}A'(\Lambda^*(\theta))}{V^*_g(\theta)}$. The total amount paid for this certificate by both good and bad providers is $\left(\phi(\theta)+\frac{1-\Lambda^*(\theta)}{\Lambda^*(\theta)}\right)V^*_g(\theta)$. On the margin, the lost revenue from a reduction of quality is given by the product of these two terms which is independent of the revenue from good providers. As a result, the comparative static does not depend on the cost of targeting views $c$. Intuitively, it is worthwhile to assign more untargeted bad provider views if they are relatively cheaper, since targeted views to good providers involve an additional cost for the platform.
	
	In particular, Proposition \ref{prop:comp_statics_gamma} and its proof highlight that there are model parameters such that $\Lambda^*(\overline{\theta})=1$ for a given $\gamma$ but $\Lambda^*(\overline{\theta})<1$ for some $\gamma'<\gamma$. So falling ad revenue can result in platforms abandoning perfect certification, as was perhaps the case for Twitter.
	
	\paragraph{\emph{Costs of targeted views}}
	We conduct a similar comparative static for the cost $c$ of directing interested views at good providers. Again, there is a natural interpretation: more information on viewers, improved algorithms and analytics reduce costs of targeting interested viewers. To consider the effect of such changes, we introduce a parameter $\kappa >0$ (that only appears in the following discussion) such that the cost of $v_g$ targeted views to good providers is $\kappa c(v_g)$.
	
	\begin{restatable}{proposition}{compstaticskappa}
		\label{prop:comp_statics_kappa}
		Let the cost $\kappa c(v_g)$ of interested views $v_g$ be parametrized by $\kappa> 0$. Then, the optimal mechanism $(V^*_g,\Lambda^*)$ satisfies the following properties.
		\begin{enumerate}
			\item[(i)] For every $\theta\in\Theta$, the quality $\Lambda^*(\theta)$ is the same for all $\kappa>0$.
			\item[(ii)] For every $\theta\in\Theta$, the quantity of views $V^*_g(\theta)$ is nonincreasing in $\kappa$.
			\item[(iii)] The set of good provider values $\{\theta\in \Theta\;|\; V^*_g(\theta)>0\}$ that are served is the same for all $\kappa>0$.
		\end{enumerate}
	\end{restatable}
	
	
	Proposition \ref{prop:comp_statics_kappa} argues that when targeting improves (that is, $\kappa$ falls) so that it becomes cheaper to find engaged consumers for good content providers, the quality of certificates do not change. Instead, good content providers enjoy more views and the platform earns more revenue from bad providers in proportion. This comparative static is immediate from the expressions for the optimal mechanism in Proposition \ref{prop:opt_mech}.
	
	
	\paragraph{\emph{Shape of consumer attention function}} \label{prop:comp_statics_shape}
	Lastly, we also examine the effect of making the attention function $A$ more concave or convex. As is clear from the first-order condition \eqref{eq:FOCmu}, the shape of the attention function determines how the platform trades off lower certificate quality against higher revenue from bad providers. Specifically, if the platform optimally chooses perfect certification for a value $\theta$ and a given attention function, it will also be optimal for the platform to choose perfect certification at this value for any other attention function that has higher slope $A'(1)$ at quality 1. Simply put, a higher slope implies that it is more costly---in terms of lost attention---for the platform to generate revenue from bad providers.
	
	The concavity or convexity of attention reflect consumer preferences. The attention function can be interpreted as the costs of reading content but, in a more flexible formulation, one might imagine this also reflecting anticipated benefits from good content relative to the costs of bad content. Since the attention function satisfies $A(0)=0$ and $A(1)=1$, it lies above the 45 degree line when it is concave and  and below when it is convex. Concavity is consistent with consumers who are particularly keen to find good content and suffer relatively little from the inconvenience associated with looking at some bad content. Instead, those with convex attention can be understood as being harmed by even a little bad content. Different kinds of media content might be thought of as differing on this scale: scrolling past bad entertainment content is perhaps only an inconvenience whereas consuming fake news is more harmful. To the extent that concavity or convexity of the attention function captures features of these different kinds of social media, the following result suggests that one would observe a greater amount of bad content on entertainment-oriented social media rather than news-related media.\footnote{Of course, this is a \textit{ceteris paribus} statement and one might expect, for example, that bad providers value views differently across these different kind of media.} 
	
	\begin{restatable}{proposition}{convexity}
		\label{prop:convexity}
		Suppose the attention function $A(\hat{\lambda};\hat{\alpha})=A(\hat{\lambda})^{\hat{\alpha}}$ is parametrized by a constant $\hat{\alpha} > 0$. Denote the optimal mechanism for a given $\hat{\alpha}$ as $(V^*_{g,\hat{\alpha}},\Lambda^*_{\hat{\alpha}})$.  Then for every $\theta\in\Theta$ and every $0<\alpha'<\alpha$ with $V^*_{g,\alpha}(\theta)>0$, we have $V^*_{g,\alpha'}(\theta)>0$ and $\Lambda^*_{\alpha}(\theta)\geq \Lambda^*_{\alpha'}(\theta)$.
		
	\end{restatable}
	
	Note that the attention function in the statement of the proposition is arbitrary so there need not be a unique $\lambda$ that maximizes $R(\phi(\theta),\lambda)$ for all $\theta\in\Theta$. Nonetheless, we obtain the requisite monotonicity for $\Lambda^*(\theta)$ (which recall, is defined to be the highest $\lambda$ that maximizes $R(\phi(\theta),\lambda)$.\footnote{Proposition \ref{prop:1proof} in the appendix provides a more general treatment of comparative statics under parameterized transformations of $A$.} More generally, the result extends to attention functions and concave transformations for which there is a unique $\lambda$ that maximizes $R(\phi(\theta),\lambda)$ for all $\theta\in \Theta$ under both the original attention function and the concave transformation.
	

	
	For intuition, consider a $\theta$ such that $\Lambda^*(\theta)\in (0,1)$ is interior. Recall that $\Lambda^*(\theta)$ must satisfy the first-order condition
	$$\left(\phi(\theta)+\frac{1-\Lambda^*(\theta)}{\Lambda^*(\theta)}\right)A'(\Lambda^*(\theta)) - \frac{A(\Lambda^*(\theta))}{\Lambda^*(\theta)^{2}} +\frac{\gamma}{\Lambda^*(\theta)^2}=0.$$
	Take an increasing, differentiable, strictly concave function $\mu$ such that $\mu(0)=0$ and $\mu(1)=1$. Replacing $A$ by the concave transformation $\mu\circ A$ on the left side of the first-order condition, we get
	$$\left(\phi(\theta)+\frac{1-\Lambda^*(\theta)}{\Lambda^*(\theta)}\right)A'(\Lambda^*(\theta))\mu'(A(\Lambda^*(\theta)) - \frac{A(\Lambda^*(\theta))}{\Lambda^*(\theta)^{2}}\frac{\mu(A(\Lambda^*(\theta))}{A(\Lambda^*(\theta))} +\frac{\gamma}{\Lambda^*(\theta)^2}<0.$$
	This implies that, for the attention function $\mu\circ A$, there are values of $\lambda<\Lambda^*(\theta)$ such that $R(\phi(\theta),\lambda)>R(\phi(\theta),\Lambda^*(\theta))$. To understand the inequality, observe that, relative to the original first-order condition, the first term is scaled by the marginal $\mu'(A(\Lambda^*(\theta))$ and the second term by the average $\frac{\mu(A(\Lambda^*(\theta))}{A(\Lambda^*(\theta))}$. But for strictly concave functions from the unit interval to the unit interval, the average $\frac{\mu(a)}{a}$ is bigger than 1 and $\frac{\mu(a)}{a}> \mu'(a)$; that is, the average is greater than the marginal for all $a\in (0,1)$. The intuition is similar to that of the comparative static for the cost $\gamma$: the concave transformation scales the benefits of bad provider views more than proportionally to its cost, and so the platform generates more revenue from bad providers.
	

	\subsection{Comparison to perfect certification}\label{subsec:compare_to_perfect}
	
	Since regulators have suggested that consumers would benefit from enforced perfect certification, it is useful to compare the optimal mechanism with the one studied in the perfect certification benchmark of Section \ref{subsec:simple}. Before we state the comparison, recall that, when perfect certification $\lambda=1$ is enforced, the optimal number of targeted views is given by $V^{s,1}_{g}(\theta)=c^{-1}(\max\{\phi(\theta)-\gamma,0\})$. In other words, the platform only serves values $\theta\in\Theta$ such that their virtual values satisfy $\phi(\theta)>\gamma$.
	
	\begin{restatable}{proposition}{welfaresummary}
		\label{prop:welfaresummary}
		The optimal mechanism $(V^*_g,\Lambda^*)$ must satisfy one of the following two properties.
		\begin{enumerate}
			\item Perfect certification is optimal: $\Lambda^*(\theta)=1$ for all $\theta\in \Theta$ and consequently, $V^*_{g}(\theta)>0$ if, and only if, $\phi(\theta) > \gamma$.
			\item Enforced perfect certification reduces content diversity: $\Lambda^*(\theta)<1$ for some $\theta\in \Theta$ with $\phi(\theta)>\gamma$ and $V^*_{g}(\theta)>0$
			for some $\theta\in \Theta$ with $\phi(\theta)<\gamma$.
		\end{enumerate}
	\end{restatable}
	
	
	We already argued (after the statement of Proposition \ref{prop:opt_mech}) that the optimal mechanism features more content diversity than the single or two certificate benchmarks. What Proposition \ref{prop:welfaresummary} additionally shows is that, when perfect certification is not optimal, the platform serves \textit{strictly} more values and there is a quality, content-diversity tradeoff. In other words, the optimal mechanism never takes the form that $\Lambda^*(\theta)=1$ for all $\theta\in \Theta$ with $\phi(\theta)>\gamma$ when $V^*_g(\theta)>0$ for some $\theta\in \Theta$ with $\phi(\theta)< \gamma$.
	
	This result suggests that the total engagement $\int_{\theta} A(\Lambda(\theta))V_g(\theta)f(\theta)d\theta$ (our measure of consumer welfare) may be higher or lower with enforced perfect certification depending on whether the value distribution $F$ assigns a greater mass to values above or below $\phi^{-1}(\gamma)$. But note that additionally, for all values $\theta\in \Theta$  with $\phi(\theta)>\gamma$ to which the optimal mechanism assigns a lower quality $\Lambda^*(\theta)<1$, the number of good provider views $V^*_g(\theta)> V^{s,1}(\theta)$ are strictly higher. This is because, by definition (from Proposition \ref{prop:opt_mech}) if $\Lambda^*(\theta)<1$, it implies that
	$$\left[\phi(\theta)+\frac{1-\Lambda^*(\theta)}{\Lambda^*(\theta)}\right]A(\Lambda^*(\theta))-\frac{\gamma}{\Lambda^*(\theta)} >  \phi(\theta)-\gamma$$
	and consequently
	$$V^*_g(\theta)=c'^{-1}\left(\left[\phi(\theta)+\frac{1-\Lambda^*(\theta)}{\Lambda^*(\theta)}\right]A(\Lambda^*(\theta))-\frac{\gamma}{\Lambda^*(\theta)}\right) >   c'^{-1}\left(\phi(\theta)-\gamma\right)=V^{s,1}_g(\theta).$$
	Indeed for $\theta$ such that $\phi(\theta)=\gamma$, the number of views under enforced perfect certification $V^{s,1}_g(\theta)=0$ whereas $V^*_g(\theta)>0$ and so engagement is higher in a neighborhood of this $\theta$ when the platform is allowed to sell imperfect certification. Put differently, even for values in the region $\theta\in \Theta$  with $\phi(\theta)>\gamma$, the effect of enforced perfect certification on engagement is ambiguous.

	\subsubsection*{Linear Attention}
	
\begin{figure}[h]
		\begin{center}
			\pgfplotsset{scaled y ticks=false} \begin{tikzpicture}     \begin{axis}[width=10cm, height=8cm, xlabel={$\phi(\theta)$},  xmin=0, xmax=1,         ymin=-0.03, ymax=0.06,         xtick={0,0.25,0.5,0.75,1}, every axis y label/.style={at={(current axis.north west)},above=2mm},  ylabel={$A(\Lambda^*(\theta))V^*_g(\theta)-V^{s,1}_g(\theta)$},       ytick={}, yticklabel={\empty},        grid=both,         major grid style={line width=.2pt,draw=gray!50},         axis lines=middle,         enlargelimits,         legend pos=south east     ]     \addplot[         domain=0:0.75,          samples=400,          thick,         blue,     ] {0.25*(1-(1-x)^0.5)/(1-x)^0.5-max((x-0.25)/2,0)} ;   \addplot[         domain=0.75:1,          samples=400,          thick,         blue,     ] {0};    \end{axis} \end{tikzpicture}
		\end{center}%
		\caption{Difference in engagement between the optimal mechanism and enforced perfect certification. Parameters:  $A(\lambda)=\lambda,\gamma=1/4$, $c(v_g)=v^2_g/2$.}
		\label{fig:opt_vs_perfect}
	\end{figure}
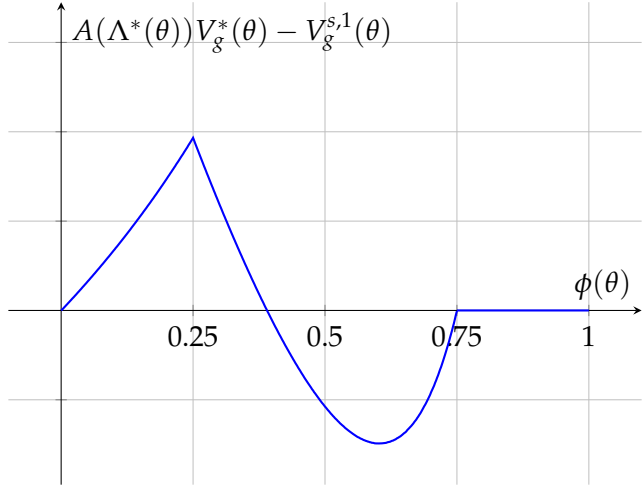

	We now return to the linear case to study this tradeoff more explicitly and to examine the effect of enforced perfect certification on total engagement. Recall that, with a linear attention function, the quality of certification (see expression \eqref{eq:linear_opt}) in the optimal mechanism is perfect ($\Lambda^*(\theta)=1$) for $\theta\in \Theta$ such that $\phi(\theta)>1-\gamma$. For these $\theta$, there will no difference between the levels of engagement ($V^*_g(\theta)$) from the optimal mechanism and that ($V^{s,1}_g(\theta)$) from enforced perfect certification. For $\theta\in \Theta$ such that $1-\frac{1}{4\gamma}<\phi(\theta)<\gamma$, the platform assigns zero views ($V^{s,1}_g(\theta)=0$) under enforced perfect certification but positive views ($V^*_g(\theta)>0$) in the optimal mechanism. Clearly for these values, engagement is higher in the optimal mechanism. For intermediate values $\gamma<\phi(\theta)<1-\gamma$ there are offsetting effects.\footnote{Recall that we are focusing on the nontrivial case of $\gamma<1/2$.} For example, when $\theta$ is just lower than $1-\gamma$, the total number of views is higher in the optimal mechanism but this is more than offset by the drop in quality. For these values, engagement is higher under enforced perfect certification. Figure \ref{fig:opt_vs_perfect} shows the difference in levels of engagement as a function of the virtual value for the parameters of our running example.
	
	Which scenario results in higher total engagement depends on the value distribution $F$. Loosely speaking, if $F$ is concentrated on values of $\theta\in\Theta$ for which $A(\Lambda^*(\theta))V^*_g(\theta)-V^{s,1}_g(\theta)>0$, then enforced perfect certification leads to lower total engagement and vice versa. It is easy to construct examples of distributions $F$ such that either case arises.
	
	\subsubsection*{Small cost $\gamma$ of untargeted views}
	
	It is possible to characterize when enforced perfect certification is and is not beneficial in the special case where, relative to the cost of targeted views $c(v_g)$, the cost $\gamma$ of untargeted views becomes small.\footnote{Stated in terms of the probability $\rho$ of finding an interested user with an untargeted view described in footnote \ref{footnote_rho}, we are effectively assuming that the ratio $\frac{\gamma}{\rho}$ is not falling, so that targeting is still necessary for good providers.} This special case is particularly relevant as there are many platforms whose opportunity cost, in terms of lost advertising revenue per view (our preferred interpretation of $\gamma$), is small.
	
	A natural starting point is to evaluate the linear attention case ($A(\lambda)=\lambda$) for which the optimal mechanism is characterized in expression \eqref{eq:linear_opt}. For $\theta\in\Theta$ such that $\phi(\theta) \ge 1$, optimal certification $\Lambda^*(\theta)=1$ is perfect regardless of the value of $\gamma$. Since $\Lambda^*(\theta)=\sqrt{\frac{\gamma}{1-\phi(\theta)}}$ for $\theta\in\Theta$ such that $\phi(\theta)<1-\gamma$, this implies that the optimal certification $\Lambda^*(\theta)\to 0$ as $\gamma\to 0$ for all $\theta$ such that $\phi(\theta)<1$. Moreover, the optimal number of views $V^*_{g}(\theta)=	c'^{-1}\left(\phi(\theta)\Lambda^*(\theta)+1-\Lambda^*(\theta)-\frac{\gamma}{\Lambda^*(\theta)}\right) \to c'^{-1}(1)$ as $\gamma\to 0$ for all $\theta$ such that $\phi(\theta)<1$. Therefore, under the optimal mechanism, the total engagement for values $\theta$ such that $\phi(\theta)<1$ vanishes as $\gamma$ becomes small. By contrast, the platform provides views $V^{s,1}_{g}(\theta)=c'^{-1}\left(\max\{\phi(\theta)-\gamma,0\}\right)$ for all $\theta\in \Theta$ under enforced perfect certification. That is, there is positive total engagement for values $\theta$ such that $0<\phi(\theta)<1$ even when $\gamma$ is small. This implies that, for small $\gamma$ and linear attention, enforced perfect certification unambiguously leads to higher total engagement.
	
	The next result shows that, if we assume that the attention function and the cost of targeted views are power functions, then enforced perfect certification unambiguously improves total engagement only if the attention function is not very concave. 
	
	\begin{restatable}{proposition}{polynomial}
		\label{prop:polynomial}
		Suppose $A(\lambda)=\lambda^{\alpha}$ where $\alpha\in(0,1)$ and $c(v_g)=\frac{v_g^{\sigma}}{\sigma}$ where $\sigma>1$. Then, for all $\theta\in\Theta$, the optimal mechanism $(V^*_{g},\Lambda^*)$	satisfies 
		\begin{align*}
			& \lim_{\gamma\downarrow 0}\Lambda^{*}(\theta)=0 \;\;\text{ and } \\
			& \lim_{\gamma\downarrow 0}A(\Lambda^{*}(\theta))V^*_{g}(\theta)=\begin{cases}
				0 & \text{if }\alpha\in \left(\frac{1}{\sigma},1\right),\\
				\alpha^{\frac{1}{\sigma-1}} & \text{if }\alpha=\frac{1}{\sigma},\\
				\infty & \text{if }\alpha\in \left(0,\frac{1}{\sigma}\right).
			\end{cases}
		\end{align*}
		Consequently, there exists a $\underline{\gamma}>0$, such that for all $\gamma<\underline{\gamma}$ enforced perfect certification leads to higher (lower) total engagement than the optimal mechanism when $\alpha< (>)\frac{1}{\sigma}$.
	\end{restatable}
	
	
	Proposition \ref{prop:polynomial} characterizes two properties of the optimal mechanism when $\gamma$ is small. The first is that, when the cost $\gamma$ of untargeted views becomes small, the platform provides more such views to bad providers and this results in the quality degrading for all certificates. This property might seem intuitive but it relies on the fact that the attention function is concave; in particular, it is not true for the linear attention case we discussed above or the convex attention case we discuss below. But this does not imply that engagement also becomes small since the platform may find it optimal to increase the number of good provider views at a faster rate. Engagement grows unboundedly only when the attention function is sufficiently concave. Intuitively, for small $\alpha$, the attention function $A(\lambda)=\lambda^{\alpha}$ is very steep for $\lambda$ close to zero and consequently, if the platform does not increase the number of engaged views for good providers in combination with the untargeted views for bad providers, this results in a large loss of attention and therefore revenue. 
	
	Proposition \ref{prop:polynomial} has two implications. For less concave attention functions, enforced perfect certification is essential since the total engagement goes to zero if the platform is allowed to degrade certificate quality. Conversely, for more concave attention functions, enforced perfect certification can result is large engagement losses because, even though the platform degrades certificate quality, they are willing to spend large amounts on targeted views of good provider content to generate engagement.
	
	
	The above result restricts attention to concave attention functions so it is natural to ask whether its conclusions extend to convex attention functions. They do not. To see this, recall the first-order condition
	$$\Lambda^*(\theta)^{\alpha}\left[1-\alpha-(\phi(\theta)-1)\alpha\Lambda^*(\theta) \right]=\gamma$$
	that $\Lambda^*(\theta)$ satisfies if it is interior (to see this, substitute $A(\lambda)=\lambda^{\alpha}$ into Equation \eqref{eq:FOCmu}). Consider $\theta\in\Theta$ such that $\phi(\theta)=0$. When $\gamma$ is small, the optimal $\Lambda^*(\theta)$ is interior and so the first-order condition holds. This in turn implies that the term in the square brackets in the above expression must be positive so
	$$1-\alpha+\alpha\Lambda^*(\theta) \geq 0 \iff \Lambda^*(\theta)\geq 1-\frac{1}{\alpha}.$$ Thus, as we take $\gamma\downarrow 0$, the quality $\Lambda^*(\theta)$ is bounded above zero so does not converge to zero as it does with concave attention shown in Proposition \ref{prop:polynomial}. Moreover, for this $\theta$, the number of views $V^*_g(\theta)$ is also bounded above zero. This will also be true for $\theta$ such that $\phi(\theta)$ is negative but close to zero. Thus, for these types, the optimal mechanism yields strictly positive engagement for small $\gamma$ but such types are unserved under enforced perfect certification. In other words, case 2 of Proposition \ref{prop:welfaresummary} applies: for some values of $\theta$, enforced perfect certification may be beneficial for engagement, but it also reduces content diversity in a way that lowers engagement for other $\theta$.\footnote{When $\alpha$ grows large, the optimal mechanism converges to the single certificate optimum with perfect certification.}  Thus, when the attention function is convex, whether or not enforced perfect certification leads to greater total engagement depends on the weights the distribution $F$ places on different values $\theta$.
	
	
	
	\section{Extension: Addiction and Losses from Bad Content}\label{subset:losses}
	
	In this section, we adapt our model in two ways in order to assess two policy-relevant concerns. First that viewing bad content may be costly for consumers, and second that consumers may suffer from digital addiction. We separately implement both into the model in rather straightforward ways and find intuitive results. As bad content becomes more costly, platforms implement cleaner certification; and as consumers are more addicted, certification becomes worse and consumers are exposed to more bad content. We then briefly revisit some of the results in the body of the paper in light of these modifications.
	
	Recall the foundation for the attention function that we described in Section \ref{sec:model}. The consumer reads a given piece of content if $\lambda-q>0$ where $\lambda$ is the belief that the content is good (the consumer values good content at one and bad content at zero) and $q$ is the cost of reading. We had assumed that the cost $q$ is distributed according to $A$ with $A(0)=0$ and $A(1)=1$. 
	
	First, suppose reading bad content generates losses $b>0$ for consumers rather than zero. In this case, a consumer reads a given piece of content when 
	$$\lambda-(1-\lambda)b-q>0 \iff q<\lambda-(1-\lambda)b.$$
	In effect, this transforms the attention function to become $A_b(\lambda):=A(\max\{\lambda(1+b)-b,0\})$; where in this expression, the maximum simply ensures that the argument of the attention function lies in the domain $[0,1]$. Note that the function $A_b(\lambda)$ takes value zero for $\lambda\leq \frac{b}{1+b}$ but it is strictly for $\lambda> \frac{b}{1+b}$.
	
	We additionally consider the second possibility that consumers are addicted and experience a loss $z\in (0,A^{-1}(\gamma))$ if they do not read a piece of content. Then, they read if $$\lambda-q>-z \iff q<\lambda+z.$$ In this case, addiction is equivalent to transforming the attention function to $A_z(\lambda):=A(\min\{\lambda+z,1\})$; the minimum here serves the same purpose as the maximum does for $A_b$—ensuring that the argument lies in $[0,1]$. $A_z(\lambda)$ is strictly increasing for $\lambda< 1-z$ and takes value one when $\lambda \geq 1-z$. We assume that $z<A^{-1}(\gamma)$, equivalently $A(z)<\gamma$, as otherwise the addition is high enough to allow the platform to make infinite profits since consumers will be willing to read content that they know is bad for sure.
	
	Note that, even though the functions $A_b$ and $A_z$ are nondecreasing rather than strictly increasing (an assumption we made for the attention function $A$), the mechanism defined in Proposition \ref{prop:opt_mech} is optimal in both cases. We thus use similar notation $(V^*_{g,b},\Lambda^*_{b})$ and $(V^*_{g,z},\Lambda^*_{z})$ respectively to denote the optimal mechanisms in these cases.
	
	The following result shows that platform content is of higher quality when costs $b$ of bad content are higher or consumers have a lower level $z$ of addiction. Note that the condition that $\frac{1}{A(\lambda)}\frac{\partial A(\lambda)}{\partial \lambda}$ is stictly decreasing is satisfied if $A(\lambda)=\lambda^{\alpha}$ for any $\alpha>0$. \footnote{As with Proposition \ref{prop:convexity}, although this condition does not imply that there is a unique optimal mechanism, the condition can be weakened if a unique optimum exists.}
	
	\begin{restatable}{proposition}{losses}
		\label{prop:losses}
		Let $A_{\hat{b}}(\lambda)=A(\max\{\lambda(1+\hat{b})-\hat{b},0\})$ with $\hat{b}>0$ and $A_{\hat{z}}(\lambda)=A(\min\{\lambda+z,1\})$ with $0<\hat{z}<A^{-1}(\gamma)$ where $A(\lambda)$ is an attention function such that $\frac{1}{A(\lambda)}\frac{\partial A(\lambda)}{\partial \lambda}$ is strictly decreasing.
		
		Then, for every $\theta\in\Theta$ and every $0<b'<b$ with $V^*_{g,b}(\theta)>0$, we have $V^*_{g,b'}(\theta)>0$ and $\Lambda^*_{b}(\theta)\geq \Lambda^*_{b'}(\theta)$. For every $\theta\in\Theta$ and every $0<z<z'<A^{-1}(\gamma)$ with $V^*_{g,z}(\theta)>0$, we have $V^*_{g,z'}(\theta)>0$ and $\Lambda^*_{z}(\theta)\geq \Lambda^*_{z'}(\theta)$.
		
	\end{restatable}
	
		
	
	Replacing the attention function $A$ with either $A_b$ or $A_z$ does not change any results in the paper with the exception of Proposition \ref{prop:polynomial}.\footnote{This result relies on $b=0$ allowing $\lambda$ to optimally be small.} In particular, Proposition \ref{prop:welfaresummary} holds: even if $b>0$, either enforced perfect certification does not matter, or it decreases content diversity. Greater content diversity remains beneficial for consumers since content is only read when $q<\lambda-(1-\lambda)b$ or equivalently, when the expected payoff is positive. Increasing $b$ might make it more likely that perfect certification is optimal for the platform but it does not change the conclusion that enforced perfect certification is either irrelevant, or comes with the cost of lower content diversity.\footnote{Welfare interpretations here are particularly subtle since when consumers are addicted, revealed preference arguments are more tenuous.}
	
	\section{Conclusion}
	
	We develop a simple and tractable model (generalizing \citet{mussa1978}) of steering and certification for sale that yields both positive and normative results.
	
	First, consistent with behavior in practice, a platform can benefit from imperfect certification since doing so enables the platform to earn revenues from bad content. Of course, higher profits can only be obtained if the platform carefully manages the quality of content and thereby consumer attention.
	
	Second, we demonstrate that optimal price discrimination involves offering a menu of distinct levels of certification with different quantities of views. The platform optimally offers lower-quality certificates and fewer views to good providers whose value, and therefore willingness to pay for engagement, is low. Conversely, the platform offers higher views and perfect certification to sufficiently high value good providers. The platform does not sacrifice rents from these high value providers by combining their content with that of bad providers.
	
	Third, we characterize conditions under which consumers are better off facing an unconstrained profit maximizing platform relative to one that is forced to use perfect certification. This occurs through two channels. First, some good providers with low values who do not appear under perfect certification do, in fact, garner views in the optimal mechanism. In essence, the platform subsidizes their views in order to generate revenue from bad providers. Consumers benefit from this increased content diversity. Second, in the optimal mechanism, the quantity of good provider views is less sensitive to their values relative to enforced perfect certification. In this way, the optimal mechanism is closer to the engagement maximizing benchmark in which the values of good content providers are irrelevant. Thus, in contrast to the tenor of recent policy discussions, our results highlight the welfare costs of enforced perfect certification.
	
	We examine several comparative statics of the optimal mechanism. We show that a lower ad revenue (a natural interpretation of the opportunity cost of allocating viewership) leads to a decrease of certificate quality, or, equivalently, more views being assigned to bad providers. Improved targeting raises platform profits via a higher number of views that can be provided without affecting the quality of certificates. Increasing the convexity of consumers' attention leads to higher quality certificates. Convexity captures the extent to which consumer attention drops off with a higher quantity of bad content at different quality levels.
	
	Our results speak to the ongoing discussion on content moderation in online platforms. Most obviously, our findings suggest that, in principle at least, consumers can benefit from allowing some bad content since it can be used to subsidize more good content and lead to more egalitarian content provision. It is also noteworthy, that the platform to some extent internalizes the harm associated with bad content since this content makes consumers pay less attention, and so from the platform's perspective limits its ability to raise revenue. Consequently, the platform optimally issues higher quality certificates if harms are higher. At an extreme, if the harm is sufficiently high then trivially no consumers will engage with content that may be bad and perfect certification maximizes platform profits with no need for regulatory intervention. There may be alternative reasons for regulatory intervention: most obviously, consumer protection for na\"ive consumers, and the possibility of externalities as discussed, for example in \citet{bursztyn2023product}. Our framework provides a starting point to address such questions in future work.
	
	Although we focus on the application to an online platform moderating content of good and bad type providers, there are other applications that fit our structure. One interpretation is that the bad content is any hidden advertisement the platform can introduce. One can imagine a search platform that can put hidden ads among the organic search results, and separate them from the explicit advertisements.\footnote{Although Google does not explicitly charge for organic traffic, some have argued that having ad business with Google influences organic placement. Similarly Amazon is alleged to favor suppliers that also purchase ancillary services.} Our model constructs the optimal way to mix these hidden ads into content, and highlights the potential costs of prohibiting this practice.
	
	A final interpretation is that the hidden advertisements are chosen by good content providers, but regulated by an outside force like the platform. An influencer can decide how much content to show that matches their own tastes, and therefore what their followers seek, and how much is not in their followers interest, but is subsidized by an outside advertiser. In that case, the certificate can stand in for a form of disclosure regulation: perfect disclosure regulation corresponds to announcing the type of content post by post, and may not be optimal when steering is for sale. Imperfect disclosure, as often seems to arise, can be better than perfectly enforced disclosure regulations.	
	
	\newpage
	
	\appendix
	
	\section{Omitted proofs from the text}
	This section contains the proofs to all the results from the body of the paper. For ease of reference, we restate each result prior to presenting the proof.
	
	\simplifcationlemma*
	
	\begin{proof}
		Given a mechanism $(V_g,V_b,M,P)$, we will construct another mechanism $(\tilde{V}_b,\tilde{V}_b,\tilde{M},P)$ with the desired properties stated above.
		
		First, we define $\tilde{M}(\theta)=\theta$. If $\Lambda(M(\theta))=0$, we define $$\tilde{V}_b(\theta)=V_b(\theta)$$ Conversely, if $\Lambda(M(\theta))>0$, we define
		\begin{equation}\label{eq:V'}
			\tilde{V}_b(\theta)=\frac{1-\Lambda(M(\theta))}{\Lambda(M(\theta))}V_g(\theta)\
		\end{equation}
		
		Let $\tilde{\Lambda}$ be the quality associated with mechanism $(V_g,\tilde{V}_b,\tilde{M},P)$. Observe that, by construction, 
		\begin{equation}\label{eq:mu'}
			\tilde{\Lambda}(\tilde{M}(\theta))=\Lambda(M(\theta))\qquad \text{ for all $\theta\in\Theta$.}
		\end{equation}
		
		so
		$$\theta A(\tilde{\Lambda}(\tilde{M}(\theta')))V_g(\theta') - P(\theta')=\theta A(\Lambda(M(\theta')))V_g(\theta') - P(\theta') \qquad \text{ for all } \theta,\theta' \in \Theta.$$
		In words, good providers of all values have the same payoff as the original mechanism (whether they report truthfully or misreport) and consequently $(V_g,\tilde{V}_b,\tilde{M},P)$ is incentive compatible and individually rational because the original mechanism $(V_g,V_b,M,P)$ is both.
		
		Take an $m$ in the image of $M$. If $\Lambda(m)=0$, then 
		$$\bbE[V_b(\theta)|M(\theta)=m]=\bbE[\tilde{V}_b(\theta)|M(\theta)=m]$$
		since $V_b$ and $\tilde{V}_b$ are defined to be equal for such $\theta$. When $\Lambda(m)>0$, \eqref{eq:V'} and \eqref{eq:mu'} together imply that
		\begin{align*}
			\bbE[V_b(\theta)|M(\theta)=m]  =\frac{1-\Lambda(m)}{\Lambda(m)}\bbE[V_g(\theta)|M(\theta)=m] & =\bbE\left[\frac{1-\tilde{\Lambda}(\tilde{M}(\theta))}{\tilde{\Lambda}(\tilde{M}(\theta))}V_g(\theta)\bigg|M(\theta)=m\right]\\
			& = \bbE[\tilde{V}_b(\theta)|M(\theta)=m].
		\end{align*}
		
		Consequently,
		$$\int_{\Theta} \left[ A(\Lambda(M(\theta)))V_b(\theta) - \gamma V_b(\theta) \right]f(\theta)d\theta=\int_{\Theta} \left[ A(\tilde{\Lambda}(\tilde{M}(\theta)))\tilde{V}_b(\theta) - \gamma \tilde{V}_b(\theta) \right]f(\theta)d\theta$$
		and so the platform makes the identical profit from $(V_g,V_b,M,P)$ and $(V_g,\tilde{V}_b,\tilde{M},P)$ as required.
	\end{proof}
		
	\singleone*
	
	\begin{proof}
		Since $\phi(\overline{\theta})=\overline{\theta}> \gamma$, observe that the optimal number of views $V^{s,1}_g(\theta)=c'^{-1}\left(\max\{\phi(\theta)-\gamma,0\}\right)$ for $\lambda=1$ is such that
		$$\int_0^{\overline{\theta}}  \left[ \left(\phi(\theta)-\gamma\right)V^{s,1}_g(\theta) - c(V^{s,1}_g(\theta)) \right]  f(\theta) d\theta>0.$$
		In words, the assumption $\overline{\theta}> \gamma$ ensures that the platform can make positive profits with perfect certification.
		
		The platform's profit when choosing a single certificate $\lambda$ and assigning views $V^{s,1}_g(\theta)$ is given by		
		$$\int_0^{\overline{\theta}}  \left[ \left(\phi(\theta)+\frac{1-\lambda}{\lambda}\right) A(\lambda) V^{s,1}_g(\theta) - \gamma\frac{V^{s,1}_g(\theta)}{\lambda} -c\left(V^{s,1}_g(\theta)\right)\right] f(\theta) d\theta.$$
		Differentiating this expression with respect to $\lambda$, we get
		$$\int_0^{\overline{\theta}}  \left[ \left(\phi(\theta)+\frac{1-\lambda}{\lambda}\right) A'(\lambda)  -\frac{A(\lambda)}{\lambda^2} + \frac{\gamma}{\lambda^2} \right] V^{s,1}_g(\theta) f(\theta) d\theta.$$
		which evaluated at $\lambda=1$ is 
		$$\int_0^{\overline{\theta}}  \left[ \phi(\theta) A'(1)  - 1 + \gamma \right] V_g(\theta) f(\theta) d\theta<0$$
		since $A'(1)>0$, $\phi(\overline{\theta}) A'(1) = \overline{\theta} A'(1) < 1 - \gamma$ and therefore $\phi(\theta) A'(1) < 1 - \gamma $ for all $\theta\in[0,\overline{\theta}]$ as $\phi(\theta)$ is strictly increasing. This implies that the platform can make strictly higher profits by picking a certificate $\lambda<1$ and assigning views $V^{s,1}_g(\theta)$ (a mechanism that is incentive compatible) which implies that choosing $\lambda=1$ cannot be optimal.
	\end{proof}
	
%
%
%
	
	\singletwo*
	
	\begin{proof}
		Take a $\lambda\in (0,1)$ and let $\tilde{\phi}$ be the solution to 
		$$R(\tilde{\phi},\lambda)=R(\tilde{\phi},1) \iff \tilde{\phi}-\gamma=\left(\tilde{\phi}+\frac{1-\lambda}{\lambda}\right)A(\lambda)-\gamma/\lambda.$$
		Observe that $R(\hat{\phi},1)>R(\hat{\phi},\lambda)$ for $\hat{\phi}>\tilde{\phi}$ and $R(\hat{\phi},1)<R(\hat{\phi},\lambda)$ for $\hat{\phi}<\tilde{\phi}$ because $0<A(\lambda)<1$.
		
		If $R(\tilde{\phi},\lambda)=R(\tilde{\phi},1)\leq 0$, then $R(\phi(\theta),1)>R(\phi(\theta),\lambda)$ for all $\theta\in\Theta$ such that $R(\phi(\theta),\lambda)> 0$ since this implies $\phi(\theta)>\tilde{\phi}$. Consequently, $$\Pi^s(1)=\int_0^{\overline{\theta}}\left(R(\phi(\theta),1) V^{s,1}_{g}(\theta)-c(V^{s,1}_{g}(\theta))\right)f(\theta)d\theta>\int_0^{\overline{\theta}}\left(R(\phi(\theta),\lambda) V^{s,\lambda}_{g}(\theta)-c(V^{s,\lambda}_{g}(\theta))\right)f(\theta)d\theta=\Pi^s(\lambda)$$
		since $V^{s,\hat{\lambda}}_{g}(\theta)=c'^{-1}\left(\max\{R(\phi(\theta),\hat{\lambda}),0\}\right)$ for all $\hat{\lambda}\in (0,1]$.
		
		If $R(\tilde{\phi},\lambda)=R(\tilde{\phi},1)> 0$, then $R(\phi(\theta),1)\geq 0$ for $\theta\in\Theta$ implies that $R(\phi(\theta),\lambda)>0$. Moreover, that there are $\theta\in\Theta$ such that $R(\phi(\theta),\lambda)>0>R(\phi(\theta),1)$. Consequently
		$$\{\theta\in\Theta|V^{s,\lambda}_g(\theta)>0\}\supset \{\theta\in\Theta|V^{s,1}_g(\theta)>0\}$$
		and so, in this case, content diversity is higher under $\lambda$.
	\end{proof}
	
%
%
%
%
	
	\twocertificates*
	
	\begin{proof}
		Suppose, for contradiction, that the two certificate optimum has the property $\umu^* = \omu^*$. Then, by Proposition \ref{prop:simple}, it must be the case that the quality $\umu^* = \omu^* \in(0,1)$ is interior. This implies that the derivative of the profit
		$$\int_0^{\overline{\theta}}  \left[ \left(\phi(\theta)+\frac{1-\umu^*}{\umu^*}\right) A(\lambda) V^{s,\umu^*}_g(\theta) - \gamma\frac{V^{s,\umu^*}_g(\theta)}{\umu^*} -c\left(V^{s,\umu^*}_g(\theta)\right)\right] f(\theta) d\theta$$
		with respect to $\lambda$ must satisfy
		$$\int_0^{\overline{\theta}}  \left[ \left(\phi(\theta)+\frac{1-\umu^*}{\umu^*}\right) A'(\umu^*)  -\frac{A(\umu^*)}{\umu^{*^2}} + \frac{\gamma}{\umu^{*^2}} \right] V^{s,\umu^*}_g(\theta) f(\theta) d\theta=0.$$
		Now note that $V^{s,\umu^*}_g(\theta)$ in nondecreasing in $\theta$ and that there is a $\theta'<\overline{\theta}$ such that $V^{s,\umu^*}_g(\theta)>0$ for $\theta\in [\theta',\overline{\theta}]$. This combined with the observation that the term in the square brackets above is increasing in $\theta$, implies that there exists a $\tilde{\theta} < \overline{\theta}$ such that
		$$\int_{\tilde{\theta}}^{\overline{\theta}}  \left[ \left(\phi(\theta)+\frac{1-\umu^*}{\umu^*}\right) A'(\umu^*)  -\frac{A(\umu^*)}{(\umu^*)^2} + \frac{\gamma}{(\umu^*)^2} \right] V^{s,\umu^*}_g(\theta) f(\theta) d\theta > 0.$$
		This contradicts the optimality of $\umu^* = \omu^*$ since the platform can achieve strictly higher profits by keeping the number of views as $V^{s,\umu^*}_g(\theta)$, keeping the certificate below $\tilde{\theta}$ as $\umu^*$ but increasing the quality of the certificate above $\tilde{\theta}$ (a mechanism that is incentive compatible since this change will maintain the requisite monotonicity). This completes the proof of the result.
	\end{proof}
	
	
	\optmech*
	\begin{proof}
		We maximize the objective function pointwise and show that the mechanism we obtain satisfies the necessary monotonicity properties to satisfy the incentive compatibility constraints.
		
		First, observe that, if
		\begin{equation}\label{eq:mu_ineq1}
			\left(\phi(\theta)+\frac{1-\lambda}{\lambda}\right)A(\lambda)-\frac{\gamma}{\lambda} \geq \left(\phi(\theta)+\frac{1-\lambda'}{\lambda'}\right)A(\lambda')-\frac{\gamma}{\lambda'}
		\end{equation}
		then, for any $V_g(\theta)\in\mathbb{R}_+$, the value of the objective function satisfies
		$$\left[\left(\phi(\theta)+\frac{1-\lambda}{\lambda}\right) A(\lambda)-\frac{\gamma}{\lambda}\right]V_g(\theta) - c(V_g(\theta))\geq \left[\left(\phi(\theta)+\frac{1-\lambda'}{\lambda'}\right) A(\lambda')-\frac{\gamma}{\lambda'}\right]V_g(\theta) - c(V_g(\theta))$$
		and vice versa when inequality \eqref{eq:mu_ineq1} is reversed.
		
		Consequently, there is a pointwise optimum that satisfies
		$$\Lambda^*(\theta)\in\argmax_{\lambda\in[0,1]}\left\{ \left(\phi(\theta)+\frac{1-\lambda}{\lambda}\right) A(\lambda) - \frac{\gamma}{\lambda}\right\}.$$
		We pick $\Lambda^*(\theta)$ to be the largest above maximizer. Note that
		$$\left(\phi(\theta)+\frac{1-\lambda}{\lambda}\right) A(\lambda) - \frac{\gamma}{\lambda} = (\phi(\theta)-1)A(\lambda) + \frac{A(\lambda) - \gamma}{\lambda}\;\;\longrightarrow\;\; -\infty$$
		as $\lambda\to 0$ (because $A(\lambda)\to 0$) and therefore $\Lambda^*(\theta)>0$ for all $\theta\in\Theta$.
		
		We now argue that $\Lambda^*$ is nondecreasing. By definition,
		$$\left(\phi(\theta)+\frac{1-\Lambda^*(\theta)}{\Lambda^*(\theta)}\right)A(\Lambda^*(\theta))-\frac{\gamma}{\Lambda^*(\theta)} \geq \left(\phi(\theta)+\frac{1-\lambda}{\lambda}\right)A(\lambda)-\frac{\gamma}{\lambda}$$
		for all $0<\lambda< \Lambda^*(\theta)$. Then for all $\theta'>\theta$, it must be the case that
		$$\left(\phi(\theta')+\frac{1-\Lambda^*(\theta)}{\Lambda^*(\theta)}\right)A(\Lambda^*(\theta))-\frac{\gamma}{\Lambda^*(\theta)} \geq \left(\phi(\theta')+\frac{1-\lambda}{\lambda}\right)A(\lambda)-\frac{\gamma}{\lambda}$$
		for all $0<\lambda< \Lambda^*(\theta)$ since $A(\Lambda^*(\theta))>A(\lambda)$ and $\phi$ is nondecreasing. Consequently, we must have $\Lambda^*(\theta')\geq \Lambda^*(\theta)$.
		
		
		The function $V^*_g$ as defined in the statement of the proposition is the solution to
		\begin{equation}\label{eq:Vbin1}
			V^*_g(\theta)=\argmax_{v_g\in \mathbb{R}_+}\left\{\left[\left(\phi(\theta)+\frac{1-\Lambda^*(\theta)}{\Lambda^*(\theta)}\right)A(\Lambda^*(\theta))-\frac{\gamma}{\Lambda^*(\theta)}\right]v_g - c(v_g)\right\}
		\end{equation}
		and the exact expression is obtained from the first-order condition.
		
		Now observe that, because $\phi$ is nondecreasing, the function
		$$\left(\phi(\theta)+\frac{1-\Lambda^*(\theta)}{\Lambda^*(\theta)}\right)A(\Lambda^*(\theta))-\frac{\gamma}{\Lambda^*(\theta)}=\max_{\lambda\in [0,1]}\left\{ \left(\phi(\theta)+\frac{1-\lambda}{\lambda}\right) A(\lambda) - \frac{\gamma}{\lambda}\right\}$$
		is nondecreasing in $\theta$ because it is the maximum of nondecreasing functions. This, in turn, implies from \eqref{eq:Vbin1}, that $V^*$ is nondecreasing. Therefore $A(\Lambda^*(\cdot))V^*_g(\cdot)$ is nondecreasing and consequently the pointwise solution is incentive compatible as required. This completes the proof.
	\end{proof}
	
	\compstaticsgamma*
	
	\begin{proof}
		For clarity, in this proof, we use the notation $\Lambda^*_{\gamma}(\theta)$ to make the dependence on $\gamma$ explicit.
		
		Recall that $\Lambda^*_{\gamma}(\theta)>0$ for all $\theta\in\Theta$. By the definition of $\Lambda^*_{\gamma}$ from Proposition \ref{prop:opt_mech}, we have
		\begin{align*}
			& \left(\phi(\theta)+\frac{1-\Lambda^*_{\gamma}(\theta)}{\Lambda^*_{\gamma}(\theta)}\right)A(\Lambda^*_{\gamma}(\theta))-\frac{\gamma}{\Lambda^*_{\gamma}(\theta)} \geq \left(\phi(\theta)+\frac{1-\lambda}{\lambda}\right)A(\lambda)-\frac{\gamma}{\lambda} \\
			\iff & \gamma\left(\frac{1}{\lambda}-\frac{1}{\Lambda^*_{\gamma}(\theta)} \right) \geq   \left(\phi(\theta)+\frac{1-\lambda}{\lambda}\right)A(\lambda) - \left(\phi(\theta)+\frac{1-\Lambda^*_{\gamma}(\theta)}{\Lambda^*_{\gamma}(\theta)}\right)A(\Lambda^*_{\gamma}(\theta))
		\end{align*}
		for all $\lambda\in(0,\Lambda^*_{\gamma}(\theta))$.
		
		Therefore, for any $\gamma'>\gamma$, we must have
		\begin{align*}
			& \gamma'\left(\frac{1}{\lambda}-\frac{1}{\Lambda^*_{\gamma}(\theta)} \right) > \left(\phi(\theta)+\frac{1-\lambda}{\lambda}\right)A(\lambda) - \left(\phi(\theta)+\frac{1-\Lambda^*_{\gamma}(\theta)}{\Lambda^*_{\gamma}(\theta)}\right)A(\Lambda^*_{\gamma}(\theta)) \\
			\iff & \left(\phi(\theta)+\frac{1-\Lambda^*_{\gamma}(\theta)}{\Lambda^*_{\gamma}(\theta)}\right)A(\Lambda^*_{\gamma}(\theta))-\frac{\gamma'}{\Lambda^*_{\gamma}(\theta)} > \left(\phi(\theta)+\frac{1-\lambda}{\lambda}\right)A(\lambda)-\frac{\gamma'}{\lambda}
		\end{align*}
		for all $\lambda\in(0,\Lambda^*_{\gamma}(\theta))$. Therefore, we must have $\Lambda^*_{\gamma'}(\theta)\geq \Lambda^*_{\gamma}(\theta)$ as required.
	\end{proof}
	
	\compstaticskappa*
	\begin{proof}
		The first statement follows immediately from Proposition \ref{prop:opt_mech} since the cost $\kappa c(\cdot)$ does not enter the expression for $\Lambda^*$.
		
		From Proposition \ref{prop:opt_mech}, $V^*_g(\theta)$ solves
		$$V^*_g(\theta)=c^{-1}\left(\frac{1}{\kappa}\max\left\{\left[\phi(\theta)+\frac{1-\Lambda^*(\theta)}{\Lambda^*(\theta)}\right]A(\Lambda^*(\theta))-\frac{\gamma}{\Lambda^*(\theta)},0\right\}\right).$$
		The second and third statements follow immediately from this equation and properties of $c$. Note that $V^*_g(\theta)>0$ whenever the right hand side of the above equation is greater than 0 and the sign of right hand side does not depend on $\kappa$. Clearly the term in the round brackets is nonincreasing in $\kappa$ so  $V^*_g(\theta)$ must be nonincreasing in $\kappa$.
	\end{proof}
	
	\renewcommand{\theproposition}{\Alph{section}.1}
	
	The remaining comparative statics use the following result.
	
	\mattnote[inline]{The proof seems to only require the final condition when $\lambda<1$, i.e. the interior part.}
	
	\begin{proposition}\label{prop:1proof}
		Consider a family of attention functions $A(\hat{\lambda};\hat{\alpha})$ parametrized by some $\hat{\alpha}\in (\underline{\alpha},\overline{\alpha})\subset \mathbb{R}_+$ and denote the optimal mechanism for a given $\hat{\alpha}$ as $(V^*_{g,\hat{\alpha}},\Lambda^*_{\hat{\alpha}})$. Assume that $A(\hat{\lambda};\hat{\alpha})$ is (i) twice continuously differentiable, (ii) strictly decreasing in $\hat{\alpha}$ and that (iii) $\frac{1}{A(\hat{\lambda};\hat{\alpha})}\frac{\partial A(\hat{\lambda};\hat{\alpha})}{\partial \hat{\alpha}}$ is strictly increasing in $\hat{\lambda}$ whenever $\hat \lambda<1$. 
		
		Then for every $\theta\in\Theta$ and every $\alpha,\alpha'\in (\underline{\alpha},\overline{\alpha})$ with $\alpha'<\alpha$  and $V^*_{g,\alpha}(\theta)>0$, we have $V^*_{g,\alpha'}(\theta)>0$ and $\Lambda^*_{\alpha}(\theta)\geq \Lambda^*_{\alpha'}(\theta)$.
	\end{proposition}
	
	\begin{proof}
		Recall that $\Lambda^*_{\alpha}(\theta)$ is defined to be the greatest $\lambda$ that maximizes of $R_{\alpha}(\theta,\lambda)$ (like the mechanism, this notation makes the dependence on $\alpha$ explicit). Since $R_{\alpha}(\theta,\lambda)$ is continuous in $\lambda$ and in $\alpha$ (by assumption), the set of $\lambda$s that  maximize $R_{\alpha}(\theta,\lambda)$ is an upper semi-continuous correspondence in $\alpha$ for each $\theta\in\Theta$. Thus, $\Lambda^*_{\alpha}(\theta)$ is an upper semi-continuous function in $\alpha$ because it is the maximum of an upper semi-continuous correspondence.
		
		In the proof, we employ the shorthand notation
		\begin{equation}\label{eq:g_function}
			g(\hat{\alpha},\hat{\lambda}):=\left(\phi(\theta)+\frac{1-\hat{\lambda}}{\hat{\lambda}}\right)A(\hat{\lambda};\hat{\alpha}).
		\end{equation}
		Observe that
		$$\frac{\partial g(\hat{\alpha},\hat{\lambda})}{\partial \hat{\alpha}}=\frac{g(\hat{\alpha},\hat{\lambda})}{A(\hat{\lambda};\hat{\alpha})}\frac{\partial A(\hat{\lambda};\hat{\alpha})}{\partial \hat{\alpha}}.$$
		
		Consider a $\theta$ and an $\alpha$ such that $V^*_{g,\alpha}(\theta)>0$. For all $\alpha'<\alpha$, we have
		$$0<g(\alpha,\Lambda^*_{\alpha}(\theta))-\frac{\gamma}{\Lambda^*_{\alpha}(\theta)}\leq g(\alpha',\Lambda^*_{\alpha}(\theta))-\frac{\gamma}{\Lambda^*_{\alpha}(\theta)}\leq \max_{\lambda\in[0,1]} \left\{g(\alpha',\lambda)-\frac{\gamma}{\lambda}\right\}=g(\alpha',\Lambda^*_{\alpha'}(\theta))-\frac{\gamma}{\Lambda^*_{\alpha'}(\theta)}$$
		where the first inequality follows from the fact that $V^*_{g,\alpha}(\theta)>0$ and the second inequality from the fact that $g(\hat{\alpha},\hat{\lambda})$ is decreasing in $\hat{\alpha}$. Therefore, $V^*_{g,\alpha'}(\theta)>0$.
		
		Once again, consider a $\theta$ and an $\alpha$ such that $V^*_{g,\alpha}(\theta)>0$. We will argue that, there exists an $\varepsilon>0$ such that $\Lambda^*_{\alpha}(\theta)\geq \Lambda^*_{\alpha'}(\theta)$ for $\alpha'\in(\alpha-\varepsilon,\alpha)$ and $\Lambda^*_{\alpha}(\theta)\leq \Lambda^*_{\alpha'}(\theta)$ for $\alpha'\in(\alpha,\alpha+\varepsilon)$. This suffices to establish that $\Lambda^*_{\alpha}(\theta)\geq \Lambda^*_{\alpha'}(\theta)$ for all $\alpha' < \alpha$ (because $V^*_{g,\hat{\alpha}}(\theta)>0$ for all $\hat{\alpha}\in[\alpha',\alpha]$) and therefore the result.
		
		
		The proof proceeds in two steps each of which has two sub-cases.
		
		\noindent \textit{Step 1:} We first argue that there exists an $\varepsilon>0$ such that $\Lambda^*_{\alpha}(\theta)\geq \Lambda^*_{\alpha'}(\theta)$ for $\alpha'\in(\alpha-\varepsilon,\alpha)$. Obviously, this is true if $\Lambda^*_{\alpha}(\theta)=1$ so we only need to consider the case $\Lambda^*_{\alpha}(\theta)<1$. 
		
		We define the sequence
		$$\lambda_n=\sup\left\{\Lambda^*_{\alpha'}(\theta)\; \bigg|\; \alpha\left(1-\frac1n\right)\le \alpha'\le \alpha\right\}$$
		for $n=\underline{n},\underline{n}+1,\dots$ where $\underline{n}>\frac{\alpha}{\alpha-\underline{\alpha}}$ is chosen to be sufficiently large so that $\alpha\left(1-\frac1n\right)>\underline{\alpha}$. Observe that this sequence is nonincreasing and bounded so it has a limit. Observe also that, by definition,  $\lambda_n\geq \Lambda^*_{\alpha}(\theta)$ for all $n$ since the supremum is taken to include $\alpha$.
		
		We split the argument into the two possible cases.
		
		\noindent Case 1.1: $\lim_{n\to\infty} \lambda_n > \Lambda^*_{\alpha}(\theta)$.
		
		Take a quality value $\tilde{\lambda}$ satisfying $\Lambda^*_{\alpha}(\theta)<\tilde{\lambda}<\lim_{n\to\infty} \lambda_n$. For every $\varepsilon\in(0,1)$, there exists an $n>\frac{\alpha}{\varepsilon}$ such that $\lambda_n>\hat{\lambda}$. From the definition of $\lambda_n$, this implies that there exists an $\alpha'\in \left[\alpha\left(1-\frac1n\right),\alpha\right]\subseteq (\alpha-\varepsilon,\alpha]$ such that $\Lambda_{\alpha'}>\tilde{\lambda}$. This violates the upper semi-continuity of the optimal quality $\Lambda^*_{\alpha}(\theta)$ at $\alpha$. Thus, this case cannot arise.
		
		\noindent Case 1.2: $\lim_{n\to\infty} \lambda_n = \Lambda^*_{\alpha}(\theta)$.
		
		If there exists an $n$ such that $\lambda_n = \Lambda^*_{\alpha}(\theta)$, then, for any $0<\varepsilon<\frac{\alpha}{n}$ we have $\Lambda^*_{\alpha}(\theta)\geq \Lambda^*_{\alpha'}(\theta)$ for $\alpha'\in(\alpha-\varepsilon,\alpha)$ as required. Thus, for the remainder of the argument of this case, we assume that $\lambda_n>\Lambda^*_{\alpha}$ for all $n$.
		
		Take an $n\ge \underline{n}$ for which $\lambda_n<1$ and note that such an $n$ exists because $\lim_{n\to\infty} \lambda_n = \Lambda^*_{\alpha}(\theta)<1$. The cross partial at $(\alpha,\Lambda^*_{\alpha}(\theta))$ satisfies
		\begin{align*}
			\frac{\partial^2 g(\alpha,\Lambda^*_{\alpha}(\theta))}{\partial \hat{\lambda} \partial \hat{\alpha}} =&  g(\alpha,\Lambda^*_{\alpha}(\theta)) \frac{\partial}{\partial \hat{\lambda}}\left(\frac{1}{A(\Lambda^*_{\alpha}(\theta);\alpha)}\frac{\partial A(\Lambda^*_{\alpha}(\theta);\alpha)}{\partial \hat{\alpha}}\right) \\& + \frac{\partial g(\alpha,\Lambda^*_{\alpha}(\theta))}{\partial \hat{\lambda}} \left(\frac{1}{A(\Lambda^*_{\alpha}(\theta);\alpha)}\frac{\partial A(\Lambda^*_{\alpha}(\theta);\alpha)}{\partial \hat{\alpha}}\right) >0.
		\end{align*}
		The inequality follows from the facts that (i) $\frac{1}{A(\hat{\lambda};\hat{\alpha})}\frac{\partial A(\hat{\lambda};\hat{\alpha})}{\partial \hat{\alpha}}$ is strictly negative and increasing in $\hat{\lambda}$ (by assumption), (ii) $g(\alpha,\Lambda^*_{\alpha}(\theta))>0$ because $V^*_{g,\alpha}(\theta)>0$ and (iii) $\frac{\partial g(\alpha,\Lambda^*_{\alpha}(\theta))}{\partial \hat{\lambda}}<0$ because $\Lambda^*_{\alpha}(\theta)<1$ (as $\lambda_n<1$) and therefore satisfies the first-order condition
		$$ \frac{\partial g(\alpha,\Lambda^*_{\alpha}(\theta)) }{\partial \hat{\lambda}} + \frac{\gamma}{\Lambda^*_{\alpha}(\theta)^2}=0.$$ 
		Consequently, because the cross partial is continuous (by assumption), there exists an $\varepsilon>0$ such that for all $(\alpha'',\lambda'')$ in an $\varepsilon$-ball $||(\alpha'',\lambda'')-(\alpha,\Lambda^*_{\alpha}(\theta))||<\varepsilon$ around $(\alpha,\Lambda^*_{\alpha}(\theta))$, the cross-partial also satisfies $\frac{\partial^2 g(\alpha'',\lambda'')}{\partial \hat{\lambda} \partial \hat{\alpha}}>0$.
		
		Now take an $\alpha'<\alpha$ such that $||(\alpha',\Lambda^*_{\alpha'}(\theta))-(\alpha,\Lambda^*_{\alpha}(\theta))||<\varepsilon$ and $\Lambda^*_{\alpha'}(\theta)>\Lambda^*_{\alpha}(\theta)$. Such an $\alpha'$ exists because $\lim_{n\to\infty} \lambda_n = \Lambda^*_{\alpha}$ and $\lambda_n > \Lambda^*_{\alpha}$ for all $n$. For this $\alpha'$, the optimality of $\Lambda^*_{\alpha'}(\theta)$ implies that
		\begin{align*}
			& g(\alpha',\Lambda^*_{\alpha'}(\theta)) - \frac{\gamma}{\Lambda^*_{\alpha'}(\theta)} \ge g(\alpha',\Lambda^*_{\alpha}(\theta)) - \frac{\gamma}{\Lambda^*_{\alpha}(\theta)} \\
			\implies & g(\alpha',\Lambda^*_{\alpha'}(\theta))+ \int^{\alpha}_{\alpha'}\frac{\partial g(\hat{\alpha},\Lambda^*_{\alpha'}(\theta))}{\partial \hat{\alpha}} d\hat{\alpha} - \frac{\gamma}{\Lambda^*_{\alpha'}(\theta)} \ge  g(\alpha',\Lambda^*_{\alpha}(\theta))  + \int^{\alpha}_{\alpha'}\frac{\partial g(\hat{\alpha},\Lambda^*_{\alpha'}(\theta))}{\partial \hat{\alpha}} d\hat{\alpha}- \frac{\gamma}{\Lambda^*_{\alpha}(\theta)} \\
			\implies & g(\alpha',\Lambda^*_{\alpha'}(\theta))+ \int^{\alpha}_{\alpha'}\frac{\partial g(\hat{\alpha},\Lambda^*_{\alpha'}(\theta))}{\partial \hat{\alpha}} d\hat{\alpha} - \frac{\gamma}{\Lambda^*_{\alpha'}(\theta)}> g(\alpha',\Lambda^*_{\alpha}(\theta)) + \int^{\alpha}_{\alpha'}\frac{\partial g(\hat{\alpha},\Lambda^*_{\alpha}(\theta))}{\partial \hat{\alpha}} d\hat{\alpha} - \frac{\gamma}{\Lambda^*_{\alpha}(\theta)} \\
			\implies & g(\alpha,\Lambda^*_{\alpha'}(\theta)) - \frac{\gamma}{\Lambda^*_{\alpha'}(\theta)} > g(\alpha,\Lambda^*_{\alpha}(\theta)) - \frac{\gamma}{\Lambda^*_{\alpha}(\theta)}.
		\end{align*}
		The third inequality is a consequence of the fact that $\frac{\partial g(\alpha'',\Lambda^*_{\alpha'}(\theta))}{\partial \hat{\alpha}}>\frac{\partial g(\alpha'',\Lambda^*_{\alpha}(\theta))}{\partial \hat{\alpha}}$ for all $\alpha''\in[\alpha',\alpha]$ because $\frac{\partial^2 g(\alpha'',\lambda'')}{\partial \hat{\lambda} \partial \hat{\alpha}}>0$ for all $\alpha''\in[\alpha',\alpha]$ and $\lambda''\in [\Lambda^*_{\alpha}(\theta),\Lambda^*_{\alpha'}(\theta)]$ since these values of $(\alpha'',\lambda'')$ lie in the $\varepsilon$-ball around $(\alpha,\Lambda^*_{\alpha}(\theta))$.
		
		This contradicts $\Lambda^*_{\alpha}(\theta)$ being optimal at $\alpha$ and therefore it is not possible to have $\lambda_n>\Lambda^*_{\alpha}(\theta)$ for all $n$. Taken together, we have thus shown that there is an $\varepsilon>0$ such that $\Lambda^*_{\alpha}(\theta)\geq \Lambda^*_{\alpha'}(\theta)$ for $\alpha'\in(\alpha-\varepsilon,\alpha)$.
		
		\noindent \textit{Step 2:} We now argue that, there exists an $\varepsilon>0$ such that $\Lambda^*_{\alpha}(\theta)\leq \Lambda^*_{\alpha'}(\theta)$ for $\alpha'\in(\alpha,\alpha+\varepsilon)$. To do so, we employ a similar argument to the first step for the case where the optimum varies continuously, but the case where the function jumps at $\alpha$ requires an additional argument.
		
		First suppose $\Lambda^*_{\alpha}(\theta)=1$. By the optimality of $\Lambda^*_{\alpha}(\theta)=1$ for $\alpha$, we have
		$$g(\alpha,\Lambda^*_{\alpha}(\theta))-\frac{\gamma}{\Lambda^*_{\alpha}(\theta)}=\phi(\theta)-\gamma\geq g(\alpha,\Lambda^*_{\alpha'}(\theta))-\frac{\gamma}{\Lambda^*_{\alpha'}(\theta)}$$
		for all $\alpha'>\alpha$. Suppose, for contradiction, that $\Lambda^*_{\alpha'}(\theta)<1$ for some $\alpha'>\alpha$.
		
		Now $V^*_{g,\alpha}(\theta)>0$ implies that $\phi(\theta)-\gamma>0$ which, in turn, implies that $\phi(\theta)>0$. By assumption, $A(\Lambda^*_{\alpha'}(\theta);\alpha)>A(\Lambda^*_{\alpha'}(\theta);\alpha')$  and this combined with the fact that  $\phi(\theta)>0$ implies $g(\alpha,\Lambda^*_{\alpha'}(\theta))> g(\alpha',\Lambda^*_{\alpha'}(\theta))$. Consequently,
		$$\phi(\theta)-\gamma\geq g(\alpha,\Lambda^*_{\alpha'}(\theta))-\frac{\gamma}{\Lambda^*_{\alpha'}(\theta)}\implies \phi(\theta)-\gamma> g(\alpha',\Lambda^*_{\alpha'}(\theta))-\frac{\gamma}{\Lambda^*_{\alpha'}(\theta)}$$
		violating the optimality of $\Lambda^*_{\alpha'}(\theta)$. To summarize, if $\Lambda^*_{\alpha}(\theta)=1$, then we must have $\Lambda^*_{\alpha'}(\theta)=1$ for all $\alpha'>\alpha$ as required.
		
		In what follows, we consider the case of $\Lambda^*_{\alpha}(\theta)<1$. We define the sequence 
		$$\lambda_n=\inf\left\{\Lambda^*_{\alpha'}(\theta)\; \bigg|\; \alpha\le \alpha'\le \alpha\left(1+\frac1n\right)\right\}$$ for $n=\underline{n},\underline{n}+1,\dots$ where $\underline{n}>\frac{\alpha}{\overline{\alpha}-\alpha}$ is chosen to be sufficiently large so that $\alpha\left(1+\frac1n\right)<\overline{\alpha}$. Observe that this sequence is nondecreasing and bounded so it has a limit. Observe also that, by definition, $\lambda_n\leq \Lambda^*_{\alpha}(\theta)$ for all $n$.
		
		As with the previous step, we break the argument into the two possible cases.
		
		\noindent Case 2.1: $\lim_{n\to\infty} \lambda_n = \Lambda^*_{\alpha}(\theta)$.
		
		If there exists an $n$ such that $\lambda_n = \Lambda^*_{\alpha}(\theta)$, then, for any $0<\varepsilon<\frac{\alpha}{n}$ we have $\Lambda^*_{\alpha'}(\theta)\geq \Lambda^*_{\alpha}(\theta)$ for $\alpha'\in(\alpha,\alpha+\varepsilon)$ as required. Thus, for the remainder of the argument of this case, we assume that $\lambda_n<\Lambda^*_{\alpha}(\theta)$ for all $n$.
		
		As we argued in Case 1.2, when $\Lambda^*_{\alpha}(\theta)<1$, the cross partial at $(\alpha,\Lambda^*_{\alpha}(\theta))$ satisfies $\frac{\partial^2 g(\alpha,\Lambda^*_{\alpha}(\theta))}{\partial \hat{\lambda} \partial \hat{\alpha}}  >0.$ Consequently, because the cross partial is continuous (by assumption), there exists an $\varepsilon>0$ such that for all $(\alpha',\lambda')$ in an $\varepsilon$-ball $||(\alpha',\lambda')-(\alpha,\Lambda^*_{\alpha}(\theta))||<\varepsilon$ around $(\alpha,\Lambda^*_{\alpha}(\theta))$, the cross-partial also satisfies $\frac{\partial^2 g(\alpha',\lambda')}{\partial \hat{\lambda} \partial \hat{\alpha}}>0$.
		
		Now take an $\alpha'>\alpha$ such that $||(\alpha',\Lambda^*_{\alpha'}(\theta))-(\alpha,\Lambda^*_{\alpha}(\theta))||<\varepsilon$ and $\Lambda^*_{\alpha'}(\theta)<\Lambda^*_{\alpha}(\theta)$. Such an $\alpha'$ exists because $\lim_{n\to\infty} \lambda_n = \Lambda^*_{\alpha}(\theta)$ and $\lambda_n > \Lambda^*_{\alpha}(\theta)$ for all $n$. For this $\alpha'$, the optimality of $\Lambda^*_{\alpha'}(\theta)$ implies that
		\begin{align*}
			& g(\alpha',\Lambda^*_{\alpha'}(\theta)) - \frac{\gamma}{\Lambda^*_{\alpha'}(\theta)} \ge g(\alpha',\Lambda^*_{\alpha}(\theta)) - \frac{\gamma}{\Lambda^*_{\alpha}(\theta)} \\
			\implies & g(\alpha',\Lambda^*_{\alpha'}(\theta))- \int^{\alpha'}_{\alpha}\frac{\partial g(\hat{\alpha},\Lambda^*_{\alpha'}(\theta))}{\partial \hat{\alpha}} d\hat{\alpha} - \frac{\gamma}{\Lambda^*_{\alpha'}(\theta)} \ge  g(\alpha',\Lambda^*_{\alpha}(\theta))  - \int^{\alpha'}_{\alpha}\frac{\partial g(\hat{\alpha},\Lambda^*_{\alpha'}(\theta))}{\partial \hat{\alpha}} d\hat{\alpha}- \frac{\gamma}{\Lambda^*_{\alpha}(\theta)} \\
			\implies & g(\alpha',\Lambda^*_{\alpha'}(\theta))- \int^{\alpha'}_{\alpha}\frac{\partial g(\hat{\alpha},\Lambda^*_{\alpha'}(\theta))}{\partial \hat{\alpha}} d\hat{\alpha} - \frac{\gamma}{\Lambda^*_{\alpha'}(\theta)}> g(\alpha',\Lambda^*_{\alpha}(\theta)) - \int^{\alpha'}_{\alpha}\frac{\partial g(\hat{\alpha},\Lambda^*_{\alpha}(\theta))}{\partial \hat{\alpha}} d\hat{\alpha} - \frac{\gamma}{\Lambda^*_{\alpha}(\theta)} \\
			\implies & g(\alpha,\Lambda^*_{\alpha'}(\theta)) - \frac{\gamma}{\Lambda^*_{\alpha'}(\theta)} > g(\alpha,\Lambda^*_{\alpha}(\theta)) - \frac{\gamma}{\Lambda^*_{\alpha}(\theta)}.
		\end{align*}
		The third inequality is a consequence of the fact that $\frac{\partial g(\alpha'',\Lambda^*_{\alpha'}(\theta))}{\partial \hat{\alpha}}<\frac{\partial g(\alpha'',\Lambda^*_{\alpha}(\theta))}{\partial \hat{\alpha}}$ for all $\alpha''\in[\alpha',\alpha]$ because $\frac{\partial^2 g(\alpha'',\lambda'')}{\partial \hat{\lambda} \partial \hat{\alpha}}>0$ for all $\alpha''\in[\alpha,\alpha']$ and $\lambda''\in [\Lambda^*_{\alpha}(\theta),\Lambda^*_{\alpha'}(\theta)]$ since these values of $(\alpha'',\lambda'')$ lie in the $\varepsilon$-ball around $(\alpha,\Lambda^*_{\alpha}(\theta))$.
		
		This contradicts $\Lambda^*_{\alpha}(\theta)$ being optimal at $\alpha$ and thus it is not possible to have $\lim_{n\to\infty} \lambda_n = \Lambda^*_{\alpha}(\theta)$ and $\lambda_n<\Lambda^*_{\alpha}(\theta)$ for all $n$. This completes the argument for this case.
		
		\noindent Case 2.2: $\lim_{n\to\infty} \lambda_n < \Lambda^*_{\alpha}(\theta)$.
		
		Let $\tilde{\lambda}=\lim_{n\to\infty} \lambda_n < \Lambda^*_{\alpha}(\theta)$. First observe that this implies
		$$g(\alpha,\Lambda^*_{\alpha}(\theta)) - \frac{\gamma}{\Lambda^*_{\alpha}(\theta)} = g(\alpha,\tilde{\lambda})- \frac{\gamma}{\tilde{\lambda}} \implies g(\alpha,\Lambda^*_{\alpha}(\theta))  < g(\alpha,\tilde{\lambda}).$$
		The inequality follows from the fact that  $\frac{\gamma}{\tilde{\lambda}}>\frac{\gamma}{\Lambda^*_{\alpha}(\theta)}$. The equality follows from the following argument. Suppose $g(\alpha,\Lambda^*_{\alpha}(\theta)) - \frac{\gamma}{\Lambda^*_{\alpha}(\theta)} > g(\alpha,\tilde{\lambda})- \frac{\gamma}{\tilde{\lambda}}$ then, by the continuity of $g(\cdot,\cdot)$, there exists an $\varepsilon>0$ such that $g(\alpha',\Lambda^*_{\alpha}(\theta)) - \frac{\gamma}{\Lambda^*_{\alpha}(\theta)} > g(\alpha',\lambda')- \frac{\gamma}{\lambda'}$ for all $(\alpha',\lambda')$ in $||(\alpha',\lambda')-(\alpha,\tilde{\lambda})||<\varepsilon$. But this contradicts $\tilde{\lambda}=\lim_{n\to\infty} \lambda_n$.
		
		Moreover, $V^*_{g,\alpha}(\theta)>0$ implies that $g(\alpha,\Lambda^*_{\alpha}(\theta)) > 0$. Taken together, we get
		\begin{align*}
			\frac{\partial g(\alpha,\Lambda^*_{\alpha}(\theta))}{\partial \hat{\alpha}}&=\frac{1}{A(\Lambda^*_{\alpha}(\theta);\alpha)}\frac{\partial A(\Lambda^*_{\alpha}(\theta);\alpha)}{\partial \hat{\alpha}} g(\alpha,\Lambda^*_{\alpha}(\theta))> \frac{1}{A(\tilde{\lambda};\alpha)}\frac{\partial A(\tilde{\lambda};\alpha)}{\partial \hat{\alpha}} g(\alpha,\tilde{\lambda})=\frac{\partial g(\alpha,\tilde{\lambda})}{\partial \hat{\alpha}}
		\end{align*}
		because $0>\frac{1}{A(\Lambda^*_{\alpha}(\theta);\alpha)}\frac{\partial A(\Lambda^*_{\alpha}(\theta);\alpha)}{\partial \hat{\alpha}}>\frac{1}{A(\tilde{\lambda};\alpha)}\frac{\partial A(\tilde{\lambda};\alpha)}{\partial \hat{\alpha}}$ by assumption.
		
		Now, since $\frac{\partial g(\cdot,\cdot)}{\partial \hat{\alpha}}$ is continuous (by assumption), there is an $\varepsilon>0$ such that
		$$\frac{\partial g(\alpha'',\lambda'')}{\partial \hat{\alpha}}>\frac{\partial g(\alpha',\lambda')}{\partial \hat{\alpha}}$$
		for all $||(\alpha'',\lambda'')-(\alpha,\Lambda^*_{\alpha}(\theta))||<\varepsilon$ and all $||(\alpha',\lambda')-(\alpha,\tilde{\lambda})||<\varepsilon$.
		
		So take an $\alpha'>\alpha$ such that  $||(\alpha',\Lambda^*_{\alpha'}(\theta)))-(\alpha,\tilde{\lambda})||<\varepsilon$. Such an $\alpha'$ must exist since  $\tilde{\lambda}=\lim_{n\to\infty} \lambda_n$. Note that, for this $\alpha'$, we also have $||(\alpha',\Lambda^*_{\alpha}(\theta))-(\alpha,\Lambda^*_{\alpha}(\theta))||<\varepsilon$. Therefore,
		$$\frac{\partial g(\alpha',\Lambda^*_{\alpha}(\theta))}{\partial \hat{\alpha}}>\frac{\partial g(\alpha',\Lambda^*_{\alpha'}(\theta))}{\partial \hat{\alpha}}$$
		and consequently for all $\alpha''\in(\alpha,\alpha')$, we also have  
		$$\frac{\partial g(\alpha'',\Lambda^*_{\alpha}(\theta))}{\partial \hat{\alpha}}>\frac{\partial g(\alpha'',\Lambda^*_{\alpha'}(\theta))}{\partial \hat{\alpha}}.$$
		This inequality combined with the optimality of $\Lambda^*_{\alpha}(\theta)$ for $\alpha$ implies that
		\begin{align*}
			& g(\alpha,\Lambda^*_{\alpha}(\theta)) - \frac{\gamma}{\Lambda^*_{\alpha}(\theta)} \ge g(\alpha,\Lambda^*_{\alpha'}(\theta)) - \frac{\gamma}{\Lambda^*_{\alpha'}(\theta)} \\
			\implies & g(\alpha,\Lambda^*_{\alpha}(\theta)) + \int^{\alpha'}_{\alpha}\frac{\partial g(\hat{\alpha},\Lambda^*_{\alpha}(\theta))}{\partial \hat{\alpha}} d\hat{\alpha} - \frac{\gamma}{\Lambda^*_{\alpha}(\theta)} \ge  g(\alpha,\Lambda^*_{\alpha'}(\theta)) + \int^{\alpha'}_{\alpha}\frac{\partial g(\hat{\alpha},\Lambda^*_{\alpha}(\theta))}{\partial \hat{\alpha}} d\hat{\alpha}- \frac{\gamma}{\Lambda^*_{\alpha'}(\theta)} \\
			\implies & g(\alpha,\Lambda^*_{\alpha}(\theta)) + \int^{\alpha'}_{\alpha}\frac{\partial g(\hat{\alpha},\Lambda^*_{\alpha}(\theta))}{\partial \hat{\alpha}} d\hat{\alpha} - \frac{\gamma}{\Lambda^*_{\alpha}(\theta)}> g(\alpha,\Lambda^*_{\alpha'}(\theta)) + \int^{\alpha'}_{\alpha}\frac{\partial g(\hat{\alpha},\Lambda^*_{\alpha'}(\theta))}{\partial \hat{\alpha}} d\hat{\alpha} - \frac{\gamma}{\Lambda^*_{\alpha'}(\theta)} \\
			\implies & g(\alpha',\Lambda^*_{\alpha}(\theta)) - \frac{\gamma}{\Lambda^*_{\alpha}(\theta)} > g(\alpha',\Lambda^*_{\alpha'}(\theta)) - \frac{\gamma}{\Lambda^*_{\alpha'}(\theta)}.
		\end{align*} 	
		This violates the optimality of $\Lambda^*_{\alpha'}(\theta)$ for $\alpha'$ and therefore  $\lim_{n\to\infty} \lambda_n < \Lambda^*_{\alpha}(\theta)$ cannot arise. This completes the proof of the result.
	\end{proof}

	\convexity*
	\begin{proof}
		Observe that
		$$A(\hat{\lambda};\hat{\alpha})=A(\hat{\lambda})^{\hat{\alpha}} \implies \frac{\partial A(\hat{\lambda};\hat{\alpha})}{\partial \hat{\alpha}}=\ln(A(\hat{\lambda}))A(\hat{\lambda})^{\hat{\alpha}} \implies \frac{1}{A(\hat{\lambda};\hat{\alpha})}\frac{\partial A(\hat{\lambda};\hat{\alpha})}{\partial \hat{\alpha}}=\ln(A(\hat{\lambda})).$$
		This implies that the result is an immediate consequence of Proposition \ref{prop:1proof} because $A(\hat{\lambda};\hat{\alpha})$ is strictly decreasing in $\hat{\alpha}$ and $\frac{1}{A(\hat{\lambda};\hat{\alpha})}\frac{\partial A(\hat{\lambda};\hat{\alpha})}{\partial \hat{\alpha}}$ is increasing in $\hat{\lambda}$.
	\end{proof}
	
	\welfaresummary*
	\begin{proof}
		
		An exhaustive list of possibilities is Cases 1 and 2 in the statement of the Proposition plus the following two cases.
		
		\noindent Case 3: $\Lambda^{*}(\theta)=1$ for all $\theta$ such that $\phi(\theta)\ge\gamma$ and $V_{g}^{*}(\theta)>0$ for some $\theta$ such that  $\phi(\theta)\le\gamma$.
		
		\noindent Case 4:  $\Lambda^{*}(\theta)<1$ for some $\theta$ such that $\phi(\theta)>\gamma$ and $V_{g}^{*}(\theta)=0$ for all $\theta$ such that  $\phi(\theta)<\gamma$.
		
		We show that Cases 3 and 4 cannot occur. Denote the contribution to profits by the type $\theta$ under the optimal mechanism by $\pi(\theta)=R(\phi(\theta),\Lambda^*(\theta))V_{g}^{*}(\theta)-c(V_{g}^{*}(\theta))$. Similarly denote the analogous contribution to profits under perfect certification by $\pi^{s,1}(\theta)=R(\phi(\theta),1)V_{g}^{s,1}(\theta) -c(V_{g}^{s,1}(\theta) )$. These are both continuous by the theorem of the maximum. They are increasing since $R(\phi(\theta),\lambda)$ is increasing in $\theta$.
		
		Suppose Case 3 could arise. This requires that $\Lambda^{*}(\theta)=1$ for all $\theta$ with $\phi(\theta) \ge \gamma$, so $\pi(\theta)=\pi^{s,1}(\theta)=0$ for $\phi(\theta)=\gamma$, and therefore $\pi(\theta)=V_{g}^{*}(\theta)=0$ for $\phi(\theta)  \le \gamma$. This contradicts $V_{g}^{*}(\theta)>0$ for some $\theta$ such that  $\phi(\theta)\le\gamma$ as required in Case 3.
		
		Finally, suppose Case 4 could arise. This requires $\pi(\theta)=0$ for $\theta$ such that $\phi(\theta) <\gamma$, so $\pi(\theta)=0$ for $\theta$ such that $\phi(\theta)=\gamma$ by continuity. Case 4 requires further that $\lim_{\theta \downarrow \phi^{-1}(\gamma)} \Lambda^*(\theta) <1$  and $\lim_{\theta \downarrow \phi^{-1}(\gamma)} R(\phi(\theta),\Lambda^*(\theta)) =0$. To see the latter, note that, if $\lim_{\theta \downarrow \phi^{-1}(\gamma)} R(\phi(\theta),\Lambda^*(\theta)) >0$, this implies that $\lim_{\theta \downarrow \phi^{-1}(\gamma)} \pi(\theta) >0$ which violates the continuity of $\pi$. But $R(\phi^{-1}(\gamma),1)=0$ and therefore we must have $\Lambda^*(\phi^{-1}(\gamma))=1$. This provides the requisite contradiction since $\lim_{\theta \downarrow \phi^{-1}(\gamma)} \Lambda^*(\theta) <1$ and $\Lambda^*(\theta)$ is nondecreasing.
	\end{proof}
	
	\polynomial*
	
	\begin{proof}
		For clarity in the proof, we denote the optimal mechanism as $(V^*_{g,\gamma},\Lambda^*_{\gamma})$ and the true effective virtual value as $R_{\gamma}$ to make the dependence on $\gamma$ explicit.
		
		We first argue that $\lim_{\gamma\downarrow 0}\Lambda^{*}_{\gamma}(\theta)=0$ for all $\theta\in\Theta$. For contradiction, suppose there exists a $\theta\in\Theta$ and a $\lambda_l>0$ such that for every $\varepsilon>0$, there exists a $\gamma\in(0,\varepsilon)$ for which $\Lambda^{*}_{\gamma}(\theta)\geq\lambda_l$. For such a $\gamma$, observe that
		$$R_{\gamma}(\phi(\theta),\Lambda^{*}_{\gamma}(\theta))= \max_{\lambda\in [\lambda_l,1]} \left\{\left(\phi(\theta)+\frac{1-\lambda}{\lambda}\right)A(\lambda)-\frac{\gamma}{\lambda}\right\}  < \max_{\lambda\in [\lambda_l,1]} \left\{\left(\phi(\theta)+\frac{1-\lambda}{\lambda}\right)A(\lambda)\right\}$$
		and so the value of $R(\phi(\theta),\Lambda^{*}_{\gamma}(\theta))$ is bounded from above by the right term (that does not depend on $\gamma$).
		
		But suppose, instead of the optimal quality $\Lambda^*_{\gamma}$, the platform chooses quality $\lambda=\gamma$. Then, 
		$$R_{\gamma}(\phi(\theta),\gamma)=(\phi(\theta)-1)\gamma^{\alpha}+\gamma^{\alpha-1}-1.$$
		Now since $\alpha\in(0,1)$ and $\gamma<\varepsilon$, if we take $\varepsilon$ sufficiently small, then $$R_{\gamma}(\phi(\theta),\gamma)>\max_{\lambda\in [\lambda_l,1]} \left\{\left(\phi(\theta)+\frac{1-\lambda}{\lambda}\right)A(\lambda)\right\}>R_{\gamma}(\phi(\theta),\Lambda^{*}_{\gamma}(\theta))$$ because $R_{\gamma}(\phi(\theta),\Lambda^{*}_{\gamma}(\theta))$ is bounded and $\gamma^{\alpha}\to 0$, $\gamma^{\alpha-1}\to\infty$, and so $R_{\gamma}(\phi(\theta),\gamma)\to\infty$ as $\gamma\downarrow 0$. This provides the requisite contradiction and implies that $\lim_{\gamma\downarrow 0}\Lambda^{*}_{\gamma}(\theta)=0$ for all $\theta\in\Theta$.
		
		Now we derive the limit of the engagement $A(\Lambda^{*}_{\gamma}(\theta))V^*_{g,\gamma}(\theta)$ from the optimal mechanism as $\gamma\downarrow 0$. First observe that, for every $\theta\in\Theta$, the above argument implies that $R_{\gamma}(\phi(\theta),\Lambda^{*}_{\gamma}(\theta))>0$ for sufficiently small $\gamma>0$. Thus, for sufficiently small $\gamma$, the views $V^*_{g,\gamma}(\theta)>0$ and are given by the equation
		$$c'(V^*_{g,\gamma}(\theta))=R_{\gamma}(\phi(\theta),\Lambda^{*}_{\gamma}(\theta))\iff V^*_{g,\gamma}(\theta)=\left(\left(\phi(\theta)+\frac{1-\Lambda^{*}_{\gamma}(\theta)}{\Lambda^{*}_{\gamma}(\theta)}\right)\Lambda^{*}_{\gamma}(\theta)^{\alpha}-\frac{\gamma}{\Lambda^{*}_{\gamma}(\theta)}\right)^{\frac{1}{\sigma-1}}.$$
		Thus, engagement from the optimal mechanism is given by
		$$A(\Lambda^{*}_{\gamma}(\theta))V^*_{g,\gamma}(\theta)=\Lambda^{*}_{\gamma}(\theta)^{\alpha}\left(\left(\phi(\theta)+\frac{1-\Lambda^{*}_{\gamma}(\theta)}{\Lambda^{*}_{\gamma}(\theta)}\right)\Lambda^{*}_{\gamma}(\theta)^{\alpha}-\frac{\gamma}{\Lambda^{*}_{\gamma}(\theta)}\right)^{\frac{1}{\sigma-1}}.$$
		
		Since $\lim_{\gamma\downarrow 0}\Lambda^{*}_{\gamma}(\theta)=0$, the optimal quality $\Lambda^{*}_{\gamma}(\theta)$ must be less than 1 for sufficiently small $\gamma>0$. Thus, for such small values of $\gamma$, it must satisfy the first-order condition
		$$\alpha(\phi(\theta)-1)\Lambda^{*}_{\gamma}(\theta)^{\alpha-1}+(\alpha-1)\Lambda^{*}_{\gamma}(\theta)^{\alpha-2}+\frac{\gamma}{\Lambda^{*}_{\gamma}(\theta)^{2}}=0.$$
		This can be rewritten as
		$$\alpha(\phi(\theta)-1)\Lambda^{*}_{\gamma}(\theta)^{\alpha}+(\alpha-1)\Lambda^{*}_{\gamma}(\theta)^{\alpha-1}=-\frac{\gamma}{\Lambda^{*}_{\gamma}(\theta)}.$$
		We use this equation to replace the term $-\frac{\gamma}{\Lambda^{*}_{\gamma}(\theta)}$ in the expression for the engagement $A(\Lambda^{*}_{\gamma}(\theta))V^*_{g,\gamma}(\theta)$ to get
		\begin{align*}
			A(\Lambda^{*}_{\gamma}(\theta))V^*_{g,\gamma}(\theta) & =\Lambda^{*}_{\gamma}(\theta)^{\alpha}\left(\left(\phi(\theta)+\frac{1-\Lambda^{*}_{\gamma}(\theta)}{\Lambda^{*}_{\gamma}(\theta)}\right)\Lambda^{*}_{\gamma}(\theta)^{\alpha}+\alpha(\phi(\theta)-1)\Lambda^{*}_{\gamma}(\theta)^{\alpha}+(\alpha-1)\Lambda^{*}_{\gamma}(\theta)^{\alpha-1}\right)^{\frac{1}{\sigma-1}}\\
			& =\Lambda^{*}_{\gamma}(\theta)^{\alpha}\left((\phi(\theta)-1)\Lambda^{*}_{\gamma}(\theta)^{\alpha}+\Lambda^{*}_{\gamma}(\theta)^{\alpha-1}+\alpha(\phi(\theta)-1)\Lambda^{*}_{\gamma}(\theta)^{\alpha}+(\alpha-1)\Lambda^{*}_{\gamma}(\theta)^{\alpha-1}\right)^{\frac{1}{\sigma-1}}\\
			& =\Lambda^{*}_{\gamma}(\theta)^{\alpha}\left((\phi(\theta)-1)(1+\alpha)\Lambda^{*}_{\gamma}(\theta)^{\alpha}+\alpha\Lambda^{*}_{\gamma}(\theta)^{\alpha-1}\right)^{\frac{1}{\sigma-1}}\\
			& =\left((\phi(\theta)-1)(1+\alpha)\Lambda^{*}_{\gamma}(\theta)^{\sigma\alpha}+\alpha\Lambda^{*}_{\gamma}(\theta)^{\sigma\alpha-1}\right)^{\frac{1}{\sigma-1}}.
		\end{align*}
		As $\gamma\downarrow 0$, the optimal quality $\Lambda^{*}_{\gamma}(\theta)\to 0$ and therefore, the first term goes to zero, since $\alpha\sigma>0$. The second term goes to zero if $\alpha>\frac{1}{\sigma}$, to infinity	if $\alpha<\frac{1}{\sigma}$, and to $\alpha$ if $\alpha=\frac{1}{\sigma}$. This completes the proof.
	\end{proof}
	
	\losses*
	
	\begin{proof}
		To prove this result, we will invoke the argument in the proof of Proposition \ref{prop:1proof}. We cannot apply that result directly since both $A_b$ and $A_z$ do not satisfy some of the assumptions---$A(0)=0$, $A(1)=1$ and $A$ increasing---that we imposed on attention functions. That said, the argument in the proof can be applied with minor adjustments. Similar to \eqref{eq:g_function} in the proof of Proposition \ref{prop:1proof}, we employ shorthand notation
		$$g(\hat{b},\hat{\lambda}):=\left(\phi(\theta)+\frac{1-\hat{\lambda}}{\hat{\lambda}}\right)A_{\hat{b}}(\hat{\lambda})$$
		with $g(\hat{z},\hat{\lambda})$ defined analogously (with $\hat{z}$ replacing $\hat{b}$). We deliberately overload the notation for simplicity since the meaning is clear from the first argument.
		
		First, we consider the case of a $b$ and a $\theta$ such that $V^*_{g,b}(\theta)>0$. Observe that, for any $b'<b$, we have $$0<g(b,\Lambda^*_{b}(\theta))-\frac{\gamma}{\Lambda^*_{b}(\theta)}\leq g(b',\Lambda^*_{b}(\theta))-\frac{\gamma}{\Lambda^*_{b}(\theta)}\leq \max_{\lambda\in[0,1]} \left\{g(b',\lambda)-\frac{\gamma}{\lambda}\right\}=g(b',\Lambda^*_{b'}(\theta))-\frac{\gamma}{\Lambda^*_{b'}(\theta)}$$
		where the first inequality follows from the fact that $V^*_{g,b}(\theta)>0$ and the second inequality from the fact that $g(\hat{b},\hat{\lambda})$ is decreasing in $\hat{b}$. Therefore, $V^*_{g,b'}(\theta)>0$.
		
		We now argue that for all $b$, there exists an $\varepsilon>0$ such that $\Lambda^*_{b}(\theta)\geq \Lambda^*_{b'}(\theta)$ for $b'\in (b-\varepsilon,b)$ and $\Lambda^*_{b}(\theta)\leq \Lambda^*_{b'}(\theta)$ for $b'\in (b,b+\varepsilon)$. This suffices to establish the required monotonicity.
		
		First, suppose $\Lambda^*_{b}(\theta)=1$, then $\Lambda^*_{b}(\theta)\geq \Lambda^*_{b'}(\theta)$ for $b'<b$. For all $b'>b$, the optimality of $\Lambda^*_{b}(\theta)=1$ for $b$ implies that
		$$g(b,\Lambda^*_{b}(\theta))-\frac{\gamma}{\Lambda^*_{b}(\theta)}=\phi(\theta)-\gamma\geq g(b,\Lambda^*_{b'}(\theta))-\frac{\gamma}{\Lambda^*_{b'}(\theta)}.$$
		Suppose, for contradiction, that $\Lambda^*_{b'}(\theta)<1$ for some $b'>b$.
		
		Now $V^*_{g,b}(\theta)>0$ implies that $\phi(\theta)-\gamma>0$ which, in turn, implies that $\phi(\theta)>0$. By assumption, $A_b(\Lambda^*_{b'}(\theta))>A_{b'}(\Lambda^*_{b'}(\theta))$  and this combined with the fact that  $\phi(\theta)>0$ implies $g(b,\Lambda^*_{b'}(\theta))> g(b',\Lambda^*_{b'}(\theta))$. Consequently,
		$$\phi(\theta)-\gamma\geq g(b,\Lambda^*_{b'}(\theta))-\frac{\gamma}{\Lambda^*_{b'}(\theta)}\implies \phi(\theta)-\gamma> g(b',\Lambda^*_{b'}(\theta))-\frac{\gamma}{\Lambda^*_{b'}(\theta)}$$
		violating the optimality of $\Lambda^*_{b'}(\theta)$. To summarize, if $\Lambda^*_{b}(\theta)=1$, then we must have $\Lambda^*_{b'}(\theta)=1$ for all $b'>b$ as required.
		
		So consider a $b$ and a $\theta$ such that   $V^*_{g,b}(\theta)>0$ and $\Lambda^*_{b}(\theta)<1$. Since $V^*_{g,b}(\theta)>0$, we must have $\Lambda^*_{b}(\theta)>\frac{b}{1+b}$. But then, following the identical argument as in the proof of Proposition \ref{prop:1proof}, there exists an $\varepsilon>0$ such that $\Lambda^*_{b}(\theta)\geq \Lambda^*_{b'}(\theta)$ for $b'\in (b-\varepsilon,b)$ and $\Lambda^*_{b}(\theta)\leq \Lambda^*_{b'}(\theta)$ for $b'\in (b,b+\varepsilon)$. This is because there is a neighborhood around $\Lambda^*_{b}(\theta)$ such that $A_b$ satisfies all the assumptions in the statement of Proposition \ref{prop:1proof} (with $b$ playing the role of $\hat{\alpha}$). To see this, observe that $A_{b}(\lambda)$ is locally strictly decreasing in $b$ and 
		$$\frac{1}{A_b(\lambda)}\frac{\partial A_b(\lambda)}{\partial \hat{b}}=(\lambda-1)\frac{A'((1+b)\lambda- b)}{A((1+b)\lambda- b)}<0$$
		is locally increasing in $\lambda$ since $\lambda<1$ and $\frac{A'(\cdot)}{A(\cdot)}>0$ is assumed to be strictly decreasing. This completes the monotonicity argument for $b$.
		
		Now consider the case of a $z$ and a $\theta$ such that $V^*_{g,z}(\theta)>0$. We will argue that there exists an $\varepsilon>0$ such that $\Lambda^*_{z}(\theta)\leq \Lambda^*_{z'}(\theta)$ for $z'\in (z-\varepsilon,z)$ and $\Lambda^*_{z}(\theta)\geq \Lambda^*_{z'}(\theta)$ for $z'\in (z,z+\varepsilon)$. This suffices to establish the required monotonicity.		
		
		We begin by arguing that $\Lambda^*_{z}(\theta)\leq 1-z$. Suppose, for contradiction, that  $\Lambda^*_{z}(\theta)> 1-z$ and so $A_z(\Lambda^*_{z}(\theta))=1$. Then,
		$$g(z,\Lambda^*_{z}(\theta))-\frac{\gamma}{\Lambda^*_{z}(\theta)}= \phi(\theta)- 1 + \frac{1 - \gamma}{\Lambda^*_{z}(\theta)} < \phi(\theta)- 1 + \frac{1 - \gamma}{1-z}=g(z,1-z)-\frac{\gamma}{1-z}$$
		which contradicts the optimality of $\Lambda^*_{z}(\theta)$. 
		
		So first suppose that $\Lambda^*_{z}(\theta)= 1-z$. Then, for $z'>z$, we 
		$$\Lambda^*_{z}(\theta)=1-z\geq 1-z' \geq \Lambda^*_{z'}(\theta)$$
		as required. For all $z'<z$, the optimality of $\Lambda^*_{z}(\theta)=1-z$ for $z$ implies that
		$$g(z,\Lambda^*_{z}(\theta))-\frac{\gamma}{\Lambda^*_{z}(\theta)}=\phi(\theta)+\frac{z-\gamma}{1-z}\geq g(z,\Lambda^*_{z'}(\theta))-\frac{\gamma}{\Lambda^*_{z'}(\theta)}.$$
		Suppose, for contradiction, that $\Lambda^*_{z'}(\theta)<1-z$ for some $z'<z$.
		
		Now $V^*_{g,z}(\theta)>0$ implies that $\phi(\theta)+\frac{z-\gamma}{1-z}>0$ which, in turn, implies that $\phi(\theta)+\frac{z}{1-z}>0$. Moreover, $\phi(\theta)+\frac{z}{1-z}>0$ implies that $\phi(\theta)+\frac{1-\Lambda^*_{z'}(\theta)}{\Lambda^*_{z'}(\theta)}>0$ because we assumed $\Lambda^*_{z'}(\theta)<1-z$. This combined with the fact that $A_z(\Lambda^*_{z'}(\theta))>A_{z'}(\Lambda^*_{z'}(\theta))$ (by assumption) implies $g(z,\Lambda^*_{z'}(\theta))> g(z',\Lambda^*_{z'}(\theta))$. Consequently,
		$$\phi(\theta)+\frac{z-\gamma}{1-z}\geq g(z,\Lambda^*_{z'}(\theta))-\frac{\gamma}{\Lambda^*_{z'}(\theta)}\implies \phi(\theta)+\frac{z-\gamma}{1-z} > g(z',\Lambda^*_{z'}(\theta))-\frac{\gamma}{\Lambda^*_{z'}(\theta)}$$
		violating the optimality of $\Lambda^*_{z'}(\theta)$. To summarize, if $\Lambda^*_{z}(\theta)=1-z$, then we must have $\Lambda^*_{z'}(\theta) \geq 1-z$ for all $z'<z$ as required.
		
		So consider a $z$ and a $\theta$ such that $\Lambda^*_{z}(\theta)< 1-z$ and therefore $A_z(\Lambda^*_{z}(\theta))< 1$. We first note that $A_z(\Lambda^*_{z}(\theta))>0$. This is because
		$$\left(\phi(\theta)+\frac{1-\lambda}{\lambda}\right) A_z(\lambda) - \frac{\gamma}{\lambda} = (\phi(\theta)-1)A_z(\lambda) + \frac{A_z(\lambda) - \gamma}{\lambda}\;\;\longrightarrow\;\; -\infty$$
		as $\lambda\to 0$ because $A_z(0)=A(z)<\gamma$ by assumption.
		
		Therefore we are left to consider the case where $\Lambda^*_{z}(\theta)\in(0,1-z)$. Once again, for this interior case, we can use the identical argument as in the proof of Proposition \ref{prop:1proof}. To do so, we redefine $A(\lambda;\hat{\alpha})=A(\min\{\lambda+1-\hat{\alpha},1\})$ where $\hat{\alpha}\in (1-A^{-1}(\gamma),1)$. Take $\alpha=1-z$ and observe that $A(\lambda;\alpha)=A_z(\alpha)=A(\lambda+z)$; therefore $\Lambda^*_{\alpha}(\theta)=\Lambda^*_{z}(\theta)\in(0,1-z)$.	Observe that there is a neighborhood around $\Lambda^*_{\alpha}(\theta)$ such that $A(\lambda;\alpha)$ satisfies all the assumptions in the statement of Proposition \ref{prop:1proof}. To see this, observe that $A_{\alpha}(\lambda)$ is locally strictly decreasing in $\alpha$ and 
		$$\frac{1}{A_{\alpha}(\lambda)}\frac{\partial A_{\alpha}(\lambda)}{\partial \hat{\alpha}}=-\frac{A'(\lambda+1-\alpha)}{A(\lambda+1-\alpha)}<0$$
		is locally increasing in $\lambda$ since $\frac{A'(\cdot)}{A(\cdot)}>0$ is assumed to be strictly decreasing.
		
		Therefore, the argument in the proof of Proposition \ref{prop:1proof} implies that there exists an $\varepsilon'>0$ such that $\Lambda^*_{\alpha}(\theta)\geq \Lambda^*_{\alpha'}(\theta)$ for $\alpha'\in (\alpha-\varepsilon',\alpha)$ and $\Lambda^*_{\alpha}(\theta)\leq \Lambda^*_{\alpha'}(\theta)$ for $\alpha'\in (\alpha,\alpha+\varepsilon')$. Consequently, there also exists an $\varepsilon>0$ such that $\Lambda^*_{z}(\theta)\leq \Lambda^*_{z'}(\theta)$ for $z'\in (z-\varepsilon,z)$ and $\Lambda^*_{z}(\theta)\geq \Lambda^*_{z'}(\theta)$ for $z'\in (z,z+\varepsilon)$. This establishes that $\Lambda^*_{z}(\theta)\geq \Lambda^*_{z'}(\theta)$ for all $z'> z$ as required to complete the proof of this result.
	\end{proof}

	\section{The relationship between engagement and consumer welfare}\label{sec:appendixwelfare}
	
	Here we substantiate the claim in Sections \ref{sec:planner} and \ref{subsec:compare_to_perfect} that maximizing engagement is equivalent to maximizing consumer welfare.

	\setcounter{proposition}{0}
	\renewcommand{\theproposition}{\Alph{section}.\arabic{proposition}}
	
	\begin{proposition}
		Suppose $A(\lambda)=\lambda^\alpha$. Then, the consumer welfare from $v_g$ good content views assigned a certificate of quality $\lambda$ is proportional to $A(\lambda)v_g$.
	\end{proposition}
	
	\begin{proof}
		We calculate the consumer welfare from $v_g$ good content views when the content is marked with a certificate of quality $\lambda$. This, of course, implies that there are $v_b=\frac{1-\lambda}{\lambda}v_g$ bad content views assigned to the same certificate.
		
		Recall that consumers pay a cost $q$ drawn from a distribution $A$ of reading content. If content is read, the consumer receives a payoff of 1 if the content is good and 0 if it is bad. Thus, only consumers with cost $q\leq \lambda$ read the content and thus consumer welfare is given by
		\[
		v_{g}\int_{0}^{\lambda}\left(1-q\right)dA(q)-v_{b}\int_{0}^{\lambda}qdA(q).
		\]
		This expression can be rewritten as
		\begin{align*}
			v_{g}\int_{0}^{\lambda}\left(1-q\right)dA(q)-v_{b}\int_{0}^{\lambda}qdA(q) & =v_{g}\int_{0}^{\lambda}\left(\left(1-q\right)-\frac{(1-\lambda)}{\lambda}q\right)dA(q)\\
			& =v_{g}\int_{0}^{\lambda}\left(1-\frac{1}{\lambda}q\right)dA(q)\\
			& =v_{g}\left(A(\lambda)-\frac{1}{\lambda}\int_{0}^{\lambda}qdA(q)\right).
		\end{align*}
		For $A(\lambda)=\lambda^{\alpha}$, this simplifies to $\frac{1}{\alpha+1}v_{g}A(\lambda)$; that is, welfare is proportional to engagement as required.
	\end{proof}

	\section{Two certificates analysis}\label{sec:appendixtwocert}
	In this section, we provide a full characterization of the platform's optimal mechanism when restricted to two certificates.
	
	\setcounter{proposition}{0}
	
	\begin{proposition}\label{prop:twocert}
		Consider the version of the platform's problem \eqref{eq:prin_prob_simple} where $\Lambda:\Theta\to \{\umu,\omu\}$ and $V_g:\Theta\to \mathbb{R}_+$.
		
		There is an optimal mechanism $(V^{bin}_g,\Lambda^{bin})$ given by
		\begin{align*}
			& \hat{\theta}=\left\{\begin{array}{cl}
				\min\left\{\theta\in\Theta \;\bigg|\; \left(\phi(\theta)+\frac{1-\omu}{\omu}\right)A(\omu)-\frac{\gamma}{\omu}\geq \left(\phi(\theta)+\frac{1-\umu}{\umu}\right)A(\umu)-\frac{\gamma}{\umu}\right\} & \text{ if the set is non-empty,} \\
				\overline{\theta} & \text{ otherwise,} 
			\end{array}\right.\\
			& \Lambda^{bin}(\theta)=\left\{\begin{array}{cl} \omu\;\; \text{ if }\theta> \hat{\theta}\; \text{ or }\; \theta=\hat{\theta}<\overline{\theta}, \\ \umu\;\; \text{ if }\theta< \hat{\theta}\; \text{ or }\; \theta= \hat{\theta}=\overline{\theta}, \end{array}\right.\qquad \text{ and }\\
			& V^{bin}_g(\theta)=c'^{-1}\left(\max\left\{\left[\phi(\theta)+\frac{1-\Lambda^{bin}(\theta)}{\Lambda^{bin}(\theta)}\right]A(\Lambda^{bin}(\theta))-\frac{\gamma}{\Lambda^{bin}(\theta)},0\right\}\right).
		\end{align*}
	\end{proposition}
	\begin{proof}
		We maximize the objective function pointwise and show that the mechanism we obtain satisfies the necessary monotonicity properties to satisfy the incentive compatibility constraints.
		
		First, observe that, if
		\begin{equation}\label{eq:mu_ineq}
			\left(\phi(\theta)+\frac{1-\omu}{\omu}\right)A(\omu)-\frac{\gamma}{\omu} \geq \left(\phi(\theta)+\frac{1-\underline{\lambda}}{\underline{\lambda}}\right)A(\underline{\lambda})-\frac{\gamma}{\underline{\lambda}}
		\end{equation}
		then, for any $V_g(\theta)\in\mathbb{R}_+$, the value of the objective function satisfies
		$$\left[\left(\phi(\theta)+\frac{1-\omu}{\omu}\right) A(\omu)-\frac{\gamma}{\omu}\right]V_g(\theta) - c(V_g(\theta))\geq \left[\left(\phi(\theta)+\frac{1-\underline{\lambda}}{\underline{\lambda}}\right) A(\underline{\lambda})-\frac{\gamma}{\underline{\lambda}}\right]V_g(\theta) - c(V_g(\theta))$$
		and vice versa when inequality \eqref{eq:mu_ineq} is reversed.
		
		Consequently, there is a pointwise optimum that satisfies
		$$\Lambda^{bin}(\theta)\in\argmax_{\lambda\in\{\underline{\lambda},\omu\}}\left\{ \left(\phi(\theta)+\frac{1-\lambda}{\lambda}\right) A(\lambda) - \frac{\gamma}{\lambda}\right\}.$$
		For any $\theta\in \Theta$ at which both $\underline{\lambda}$ and $\omu$ are maximizers, we set $\Lambda^{bin}(\theta)=\omu$.
		
		Now note that, if \eqref{eq:mu_ineq} holds for some $\theta$, it also holds for all $\theta'>\theta$. This is because $A(\omu)>A(\underline{\lambda})$ and $\phi(\cdot)$ is nondecreasing. Consequently, $\Lambda^{bin}$ must take the cutoff form
		$$\Lambda^{bin}(\theta)=\left\{\begin{array}{cl} \omu & \text{ when } \theta> \hat{\theta},\\ \underline{\lambda} & \text{ when } \theta< \hat{\theta}.
		\end{array}\right.$$
		as in the statement of the proposition and thus $\Lambda^{bin}$ is nondecreasing.
		
		The function $V^{bin}_g$ as defined in the statement of the proposition is the solution to
		\begin{equation}\label{eq:Vbin}
			V^{bin}_g(\theta)=\argmax_{v_g\in \mathbb{R}_+}\left\{\left[\left(\phi(\theta)+\frac{1-\Lambda^{bin}(\theta)}{\Lambda^{bin}(\theta)}\right)A(\Lambda^{bin}(\theta))-\frac{\gamma}{\Lambda^{bin}(\theta)}\right]v_g - c(v_g)\right\}
		\end{equation}
		and the exact expression is obtained from the first-order condition.
		
		Now observe that, because $\phi$ is nondecreasing, the function
		$$\left(\phi(\theta)+\frac{1-\Lambda^{bin}(\theta)}{\Lambda^{bin}(\theta)}\right)A(\Lambda^{bin}(\theta))-\frac{\gamma}{\Lambda^{bin}(\theta)}=\max_{\lambda\in\{\underline{\lambda},\omu\}}\left\{ \left(\phi(\theta)+\frac{1-\lambda}{\lambda}\right) A(\lambda) - \frac{\gamma}{\lambda}\right\}$$
		is nondecreasing because it is the maximum of two nondecreasing functions. This, in turn, implies from \eqref{eq:Vbin}, that $V^{bin}$ is nondecreasing. Therefore $A(\Lambda^{bin}(\cdot))V^{bin}_g(\cdot)$ is nondecreasing and consequently the pointwise solution is incentive compatible as required. This completes the proof.
	\end{proof}

	\newpage
	
	\bibliographystyle{econometrica}
	\bibliography{SellingCheckmarks_Refs}
	
\end{document}